\documentclass[pre,amsmath,amssymb,showpacs]{revtex4}

\usepackage{graphicx}% Include figure files
\usepackage{dcolumn}% Align table columns on decimal point
\usepackage{bm}% bold math
\usepackage{epsf}
 

\begin{document}

\title{An integral equation approach to time-dependent 
kinematic dynamos in finite domains}

\author{Mingtian Xu}
\email{M.Xu@fz-rossendorf.de}
\author{Frank Stefani}
\email{F.Stefani@fz-rossendorf.de}
\author{Gunter Gerbeth}
\email{G.Gerbeth@fz-rossendorf.de}
\affiliation{Forschungszentrum Rossendorf, P.O. Box 510119,
D-01314 Dresden, Germany}
 
\date{\today}

\begin{abstract}
The homogeneous dynamo effect is at the root of cosmic 
magnetic field generation. With only a very few exceptions, 
the numerical treatment of homogeneous dynamos
is carried out in the framework  of the differential equation 
approach.
The present paper tries to facilitate the use of integral
equations in dynamo research. Apart from the pedagogical
value to illustrate dynamo action within the well-known picture of
Biot-Savart's law, the integral equation approach has a number
of practical advantages. The first advantage is its proven
numerical robustness and stability. The second and 
perhaps most important advantage is its applicability to 
dynamos in arbitrary geometries. The third advantage 
is its intimate connection to
inverse problems relevant not only for dynamos but also for
technical applications of magnetohydrodynamics.
The paper provides the first general formulation and application 
of the integral 
equation approach to time-dependent kinematic dynamos in finite 
domains. For the spherically symmetric $\alpha^2$ dynamo model 
it is shown
how the general formulation is reduced to a coupled system 
of two radial integral equations for the defining scalars of the
poloidal and the toroidal field components. 
The integral equation formulation for
spherical dynamos with general velocity fields is also derived.
Two numerical examples, the $\alpha^2$ dynamo model with radially 
varying $\alpha$, and the Bullard-Gellman model illustrate the equivalence 
of the
approach  with the usual differential equation method.

\end{abstract}
%\pacs{47.65.+a, 52.65.Kj, 91.25.Cw}  

%\tableofcontents
\maketitle

\section{Introduction}

Cosmic magnetic fields, including the fields of planets, stars, and
galaxies, result from the hydromagnetic dynamo effect \cite{KRRA,MOFF}. 
The last 
decades have seen tremendous progress
in the analytical and numerical treatment
of magnetic field generation in cosmic bodies. Recently, the 
hydromagnetic dynamo effect has been validated 
experimentally
in large liquid sodium facilities in Riga and Karlsruhe 
\cite{PRL1,MUST,PRL2,STMU,RMP}.

The usual way to treat hydromagnetic dynamos numerically is 
within the
differential equation approach. 
Supposing the fluid velocity $\bf{u}$ to be given, 
the governing differential equation is the 
induction equation for the magnetic field $\bf{B}$,
\begin{eqnarray}{\label{eq1}}
\frac{\partial {{\bf{B}}}}{\partial t}=\nabla
\times ({\bf{u}} \times {\bf{B}})
+\frac{1}{\mu_0 \sigma} \Delta {\bf{B}} \label{4} \; ,
\label{eq0}
\end{eqnarray}
with $\mu_0$ and $\sigma$ denoting the permeability of the free space 
and the electrical 
conductivity 
of the fluid, respectively. Equation (\ref{eq1}) follows directly from 
pre-Maxwell's equations and Ohm's law 
in moving conductors. Note that the
magnetic field has to be divergence-free:
\begin{eqnarray}{\label{eq2}}
\nabla \cdot {\bf{B}}=0 \label{0a} \; .
\end{eqnarray}

In case of vanishing excitations of the            
magnetic field from outside the considered finite region,          
the boundary condition for the magnetic field reads                
\begin{eqnarray}{\label{eq3}}                                                  
{\bf{B}}={\bf{O}}(r^{-3}) \; \;\; \mbox{as}                         
\; \;\; r \rightarrow \infty \label{3a} \; .                       
\end{eqnarray} 

The induction equation (1) is sufficient to treat {\it kinematic dynamo models}
in which the back-reaction of the self-excited magnetic field on the
flow is neglected. Such a simplification is justified during the
initial phase of self-excitation when the magnetic field is
weak. For stronger fields, one has to cope with {\it dynamically 
consistent dynamo models} which require the  simultaneous solution of 
the induction equation
for the magnetic field and the Navier-Stokes 
equation for the velocity. This saturation regime, however, will not be
considered in the present paper.

The spherical geometry of many cosmic bodies such as planets and stars
has simplified dynamo simulations in one important respect: 
for the spherical case the boundary 
conditions for 
the magnetic 
field can be re-formulated separately for every degree and order of the 
spherical harmonics, which  makes any particular treatment of the magnetic fields 
in the exterior superfluous.

When it comes to dynamos in other than spherical geometries,
this pleasant situation changes. Then the correct 
handling of
the non-local boundary conditions becomes non-trivial. 
Such a problem appears, e.g., in the
numerical simulation of the recent dynamo experiments
that are of cylindrical shape, but also in the simulations
of galactic dynamos. There are some ways to circumvent 
this problem: one can use simplified local boundary conditions
(``vertical field condition'') \cite{BRAN,RUZH}, one can embed the 
actual dynamo body into a sphere with the region between the actual 
dynamo  and the 
surface of the sphere virtually filled with a medium of lower 
electrical conductivity \cite{RAE1,RAE2}, or one 
can solve the Laplace equation in the exterior and fit 
the solution at the boundary to the solution in the
interior \cite{STE1}, which is, however, a tedious and time-consuming
procedure.
 
The integral equation approach that we will establish in the present 
paper is intended to change this  unsatisfactory situation 
concerning the handling
of boundary conditions.
For the steady case, and for infinite domains of 
homogeneous conductivity, 
the integral equation 
approach was 
used in a few previous papers \cite{GAI2,GAI3,FREI,DORA}.
The inclusion of boundaries, again for the steady case, was already 
delineated
in the book of Roberts \cite{ROB1}. 
In Roberts' own opinion (\cite{ROB1}, p. 74), however, 
this formulation did 
``...not appear, in general, to be very useful.'' 
In \cite{STE2} we have tried to put this pessimistic judgment
into question. In particular, we have derived from the general theory 
a system of one-dimensional integral equations for a dynamo model with a
spherically symmetric, isotropic helical turbulence parameter $\alpha$
in a finite
sphere, and we have 
re-derived analytically the solution found by Krause and 
Steenbeck \cite{KRST} for
the special case of constant $\alpha$.
In \cite{XUSG} we have investigated the 
performance of numerical schemes to solve these integral
equations for steady $\alpha^2$ dynamos. 

It should be pointed out that a similar approach had been established
earlier
under the label ``velocity-current-formulation'' 
by Meir and Schmidt \cite{MES1,MES2}. This approach had also 
the
pronounced aim to circumvent the numerical treatment outside the region of
interest.  However, the numerical focus of this work laid more on 
steady, coupled MHD problems with  small magnetic Reynolds number
than on dynamo problems.  

In the present paper, we generalize the integral equation approach
for steady dynamos in finite domains to the time-dependent
case. In contrast to earlier statements on this 
matter \cite{DORA,STE2,XUSG}  that the use 
of the Green's function of the Helmholtz equation will lead us to
a non-linear eigenvalue equation, we present here a linear formulation 
of the
eigenvalue problem. This makes the approach more attractive for numerical
treatment. 

For the paradigmatic case of a time-dependent spherically symmetric, 
isotropic  $\alpha^2$ dynamo we derive, starting from the general formulation,   
a coupled system of two
radial integral equations which is solved numerically.  The equivalence with
the results of an differential equation solver is made evident.

We  also derive the coupled system of 
radial integral equations for spherical dynamos with general velocity 
fields, and treat numerically the well-known Bullard-Gellman model 
within the new approach.

\section{General formulation}
In this section, the general form of the integral equation approach 
is developed for time-dependent dynamos 
acting in an electrically 
conducting, non-magnetic fluid  
which occupies 
a finite domain $D$ with a boundary $S$, surrounded 
by non-conducting space. Under the assumption that the velocity or
corresponding mean-field quantities are stationary, 
we generalize the basic idea and the methods from the 
steady case \cite{STE2} to the 
time-dependent case. 

We start with
pre-Maxwell's equations (the displacement current can be skipped in the 
quasistationary approximation),
\begin{eqnarray}{\label{eq4}}
\nabla \times {\bf{E}}({\bf{r}},t)&=&-\frac{\partial 
{\bf{B}}({\bf{r}},t)}{\partial t} \; , {\label{eq4a}} \\
\nabla\cdot {\bf{B}}({\bf{r}},t)&=&0\; , {\label{eq4b}} \\
\nabla\times {\bf{B}}({\bf{r}},t)&=&\mu_0{\bf{j}}({\bf{r}},t)   \; ,{\label{eq4c}}  
\end{eqnarray} 
where ${\bf{B}}({\bf{r}},t)$ is the magnetic field, ${\bf{E}}({\bf{r}},t)$ the electric 
field, ${\bf{r}}$ the position vector, $t$ the time, and $\mu_0$ the magnetic 
permeability of the 
free space. The current density ${\bf{j}}({\bf{r}},t)$ satisfies Ohm's law 
\begin{eqnarray}{\label{eq5}}
{\bf{j}}({\bf{r}},t)=\sigma({\bf{E}}({\bf{r}},t)+{\bf{F}}({\bf{r}},t))
\end{eqnarray}
inside the dynamo domain $D$, and it vanishes outside $D$.
${\bf{F}}({\bf{r}},t)$ denotes the electromotive  force (emf)
${\bf{u}}({\bf{r}})\times {\bf{B}}({\bf{r}},t)$, where ${\bf{u}}({\bf{r}})$ is
 the velocity of the fluid motion, which we suppose to be stationary.   In the
framework of mean-field electrodynamics  (see, e.g., \cite{KRRA})
${\bf{B}}({\bf{r}},t)$, ${\bf{j}}({\bf{r}},t)$  and ${\bf{F}}({\bf{r}},t)$ are
split into mean fields and fluctuating fields.  Then ${\bf{F}}({\bf{r}},t)$ 
can be fixed to the form
\begin{eqnarray}{\label{eq6}}
{\bf{F}}({\bf{r}},t)={\bf{u}}({\bf{r}})\times {\bf{B}}({\bf{r}},t)+\alpha({\bf{r}}) 
{\bf{B}}({\bf{r}},t)-\beta({\bf{r}}) \nabla\times{\bf{B}}({\bf{r}},t),
\end{eqnarray}
where  ${\bf{u}}({\bf{r}})$ and ${\bf{B}}({\bf{r}},t)$ now denote the  mean
velocity and the mean magnetic field, respectively. 
The term  $\alpha({\bf{r}}) \bf{B}({\bf{r}},t)$ describes the emf due to the
non-mirrorsymmetric  part of the turbulence ($\alpha$-effect),
and the term  $\beta({\bf{r}}) \nabla  \times {\bf{B}}({\bf{r}},t)$ describes 
another effect
which can be interpreted 
as an conductivity decrease due to the turbulence ($\beta$-effect).

The equations given so far, together with initial conditions for ${\bf{B}}$,
define an initial value problem for ${\bf{B}}({\bf{r}},t)$ which can be combined  into 
the form
of Eq. (\ref{eq0}). Together 
with the requirements that there are no surface currents
on $S$ and that ${\bf{B}}({\bf{r}},t)$ vanishes at infinity, they
allow to determine ${\bf{B}}({\bf{r}},t)$  once the concrete
dependence of ${\bf{F}}({\bf{r}},t)$ on ${\bf{B}}({\bf{r}},t)$ has been fixed. 

Under the condition 
that $\bf{u}$, $\alpha$ and $\beta$ in Eq. (\ref{eq6}) 
only depend on the position ${\bf{r}}$, the 
functions ${\bf{E}}({\bf{r}},t)$, ${\bf{B}}({\bf{r}},t)$, ${\bf{F}}({\bf{r}},t)$ 
and ${\bf{j}}({\bf{r}},t)$ may 
be written in the form:
\begin{eqnarray}{\label{eq7}}
{\bf{E}}({\bf{r}},t)={\bf{E}}({\bf{r}})e^{\lambda t},\; 
{\bf{B}}({\bf{r}},t)={\bf{B}}({\bf{r}})e^{\lambda t},\; 
{\bf{F}}({\bf{r}},t)={\bf{F}}({\bf{r}})e^{\lambda t},\;
{\bf{j}}({\bf{r}},t)={\bf{j}}({\bf{r}})e^{\lambda t},
\end{eqnarray}
where $\lambda$ is in general a complex constant.
Then Maxwell's equations and Ohm's law transform into
\begin{eqnarray}
\nabla \times {\bf{E}}&=&-\lambda {\bf{B}} \; ,{\label{eq8a}}\\
\nabla \cdot {\bf{B}}&=&0 \;, {\label{eq8b}} \\
\nabla \times {\bf{B}}&=&\mu_0 {\bf{j}} {\label{eq8c}}\; ,\\
{\bf{j}}&=&\sigma ({\bf{E}}+{\bf{F}}) \; . {\label{eq9}}
\end{eqnarray}
Hereafter, the symbols ${\bf{B}}$, ${\bf{E}}$, ${\bf{j}}$ and ${\bf{F}}$ denote the 
functions which only depend on the position ${\bf{r}}$, but not on the time $t$.

Since the magnetic field $\bf{B}$ is divergence-free, there exists a 
magnetic vector potential ${\bf{A}}({\bf{r}})$ such that
\begin{eqnarray}{\label{eq10}}
{\bf{B}}=\nabla\times {\bf{A}},
\end{eqnarray}
which implies, together with Eq. (\ref{eq8a}), that
\begin{eqnarray}{\label{eq11}}
\nabla \times( {\bf{E}}+\lambda {\bf{A}})=0 \;.
\end{eqnarray}
Therefore, ${\bf{E}}+\lambda {\bf{A}}$ has to be irrotational, and we can write 
\begin{eqnarray}{\label{eq12aa}}
{\bf{E}}+\lambda {\bf{A}}=-\nabla \varphi \; .
\end{eqnarray}
Subsequently, Ohm's law gets the form
\begin{eqnarray}{\label{eq12}}
{\bf{j}}=\sigma({\bf{F}}-\lambda {\bf{A}}-\nabla \varphi).
\end{eqnarray}

In the following, we will derive a system of integral equations that is equivalent
to the differential equation formulation.
The starting point for the first integral equation is the 
application of Biot-Savart's law on Eq. (\ref{eq8c}), leading to
\begin{eqnarray}{\label{eq12a}}
{\bf{B}}({\bf{r}})&=&\frac{\mu_0}{4 \pi} \int_D
\frac{ {\bf{j}}({\bf{r'}}) \times 
({\bf{r}}-{\bf{r'}})}{|{\bf{r}}-{\bf{r'}}|^3} \; dV' \; . 
\end{eqnarray}
It is well known that the $curl$ operator in Eq. (\ref{eq8c})
is a left inverse of the Biot-Savart operator when 
the current $\bf j$ is divergence-free and tangent to the boundary of the
domain \cite{CANT}. Both conditions are indeed fulfilled in our case.
Inserting Eq. (\ref{eq12}) into Eq. (\ref{eq12a}), 
using Eq. (\ref{eq10}) and Gauss's 
theorem, we get the first integral equation
\begin{eqnarray}{\label{eq13}}
{\bf{B}}({\bf{r}})&=&\frac{\mu_0 \sigma}{4 \pi} \int_D
\frac{ {\bf{F}}({\bf{r'}}) \times 
({\bf{r}}-{\bf{r'}})}{|{\bf{r}}-{\bf{r'}}|^3} \; dV' 
-\frac{\mu_0 \sigma \lambda}{4 \pi} \int_D
\frac{{\bf{B}}({\bf{r'}})}{|{\bf{r}}-{\bf{r'}}|} \; dV' \nonumber\\
&&+\frac{\mu_0 \sigma \lambda}{4 \pi} \int_S 
{\bf{n}} ({\bf{s'}}) \times
\frac{{\bf{A}}({\bf{s'}})}{|{\bf{r}}-{\bf{s'}}|} \; dS'
-\frac{\mu_0 \sigma}{4 \pi} \int_S \varphi({\bf{s'}}) 
{\bf{n}} ({\bf{s'}}) \times
\frac{{\bf{r}}-{\bf{s'}}}{|{\bf{r}}-{\bf{s'}}|^3} \; dS',
\end{eqnarray}
with ${\bf{n}}({\bf{s'}})$ denoting the outward directed 
unit vector 
at the boundary point $\bf{s'}$ and $dS'$ denoting an area element
at this point. 
For some purposes it might be useful to express the first 
volume integral in Eq. (\ref{eq13}) in the form
\begin{eqnarray}\label{eq13aa}
\int_D
\frac{ {\bf{F}}({\bf{r'}}) \times 
({\bf{r}}-{\bf{r'}})}   {|{\bf{r}}-{\bf{r'}}|^3} \; dV' &=&\int_D
\frac{\nabla_{r'} \times {\bf{F}}({\bf{r'}}) }{|{\bf{r}}-{\bf{r'}}|} \; dV' 
-\int_S {\bf{n}} ({\bf{s'}}) \times
\frac{{\bf{F}}({\bf{s'}})}{|{\bf{r}}-{\bf{s'}}|} \; dS' \; .
 \end{eqnarray}
For the steady case ($\lambda=0$)  Eq. (\ref{eq13})  reduces to the 
form given in \cite{STE2}, with one volume integral over $\bf{B}$ 
and one boundary integral over $\varphi$. The latter one would vanish only
in the case that the dynamo region is extended to infinity.
The time-dependence introduces now a new volume 
integral over $\bf{B}$, but also one boundary integral over the
vector potential $\bf{A}$ which cannot be reduced to a simple 
expression in $\bf{B}$. Before we focus on this point, let
us first derive the integral equation for $\varphi$.

From  Eq. (\ref{eq12}) and the demand that the current has to be 
divergence-free, $\nabla \cdot {\bf{j}}=0$, 
we get
a Poisson equation for $\varphi$:
\begin{eqnarray}{\label{eq13a}}
\Delta \varphi=\nabla \cdot ({\bf F}-\lambda {\bf A}) \; .
\end{eqnarray}
Assuming vacuum boundary condition which means that the current 
must not leave the
domain $D$, we obtain from Green's theorem the following boundary 
integral 
equation for 
$\varphi$:
\begin{eqnarray}{\label{eq14}}
p \; \varphi({\bf{r}})&=& 
\frac{1}{4 \pi} \int\limits_D  
\frac{{\bf{F}}({\bf{r'}}) \cdot ({\bf{r}}-{\bf{r'}})}{|{\bf{r}}-{\bf{r'}}|^3}
\; dV' +\frac{\lambda}{4 \pi} \int\limits_D \frac{\nabla_{r'} 
\cdot {\bf A}({\bf{r'}})}{
|{\bf{r}}-{\bf{r'}}|} \; dV'\nonumber\\
&&- \frac{\lambda}{4 \pi} \int\limits_S
{\bf{n}}({\bf{s'}}) \cdot
\frac{{\bf{A}}({\bf{s'}})}{|{\bf{r}}-{\bf{s'}}|} \; dS'
-\frac{1}{4 \pi} \int\limits_S \varphi({\bf{s'}})
{\bf{n}}({\bf{s'}}) 
\cdot \frac{{\bf{r}}-{\bf{s'}}}{{|{\bf{r}}-{\bf{s'}}|}^3} \;  dS',
\end{eqnarray}
where  $p=1$ for points $\bf{r}$ inside $D$, $p=1/2$ for 
points ${\bf{r}}={\bf{s}}$ on $S$ and $p=0$ for points $\bf{r}$
outside $D$. Again, another expression of the first volume integral
might be useful :
\begin{eqnarray}\label{eq14aa}
\int_D
\frac{ {\bf{F}}({\bf{r'}}) \cdot
({\bf{r}}-{\bf{r'}})}   {|{\bf{r}}-{\bf{r'}}|^3} \; dV' &=&-\int_D
\frac{\nabla \cdot {\bf{F}}({\bf{r'}}) }{|{\bf{r}}-{\bf{r'}}|} \; dV' 
+\int_S {\bf{n}} ({\bf{s'}}) \cdot
\frac{{\bf{F}}({\bf{s'}})}{|{\bf{r}}-{\bf{s'}}|} \; dS' \; .
\end{eqnarray}
If we use the Coulomb gauge for the vector potential, $\nabla \cdot
{\bf{A}}=0$, we
see that the second volume integral in Eq. ({\ref{eq14}}) vanishes.
Of course, one is free to use other gauges than the Coulomb
gauge. One advantage of this gauge, in view of later numerical applications, 
is that it leads to a formulation in which the vector  potential
is only needed at the boundary of the domain.

For the steady case, Eqs. (\ref{eq13}) and (\ref{eq14}) with $\lambda=0$ 
are 
sufficient to determine the magnetic field $\bf{B}$. But for the 
time-dependent case presently under consideration, 
we have to introduce the vector 
potential $\bf{A}$, at least at the
boundary, 
for completely formulating the problem. Necessarily we have to
establish another relation for $\bf{A}$ in order to make the problem 
solvable. 
From Eq. (\ref{eq10}) and Helmholtz's theorem 
(\cite{MORS}, p. 53) we can express the vector potential at 
the boundary in one of the two forms
\begin{eqnarray}{\label{eq15}}
{\bf{A}}({\bf{s}})&=&\frac{1}{4\pi}\int_D\frac{\nabla_{r'}
\times{\bf{B}}({\bf{r}}')}{|{\bf{r}}-{\bf{r}}'|}dV' 
\nonumber\\
&=&
\frac{1}{4\pi}\int_D \frac{ {\bf{B}}({\bf{r'}}) 
\times ({\bf{s}}-{\bf{r'}})}{|{\bf{s}}-{\bf{r'}}|^3} dV'
+\frac{1}{4\pi}\int_S{\bf{n}}({\bf{s'}}) \times\frac{{\bf{B}}({\bf{s'}})}
{|{\bf{s}}-{\bf{s'}}|}dS'\; .
\end{eqnarray}

The integral equations (\ref{eq13}), (\ref{eq14}) and (\ref{eq15}) 
provide another complete formulation of the problem for $\bf{B}$. 
The main advantage of this formulation is that one can avoid
any treatment of fields in the exteriour of $D$. The boundary 
conditions
are being fulfilled by solving the boundary integral
equations for $\varphi$ and $\bf{A}$.

In the next section we will sketch how the equations for $\bf{A}$ and $\varphi$ at
the boundary  can be absorbed into
a single integral equation for $\bf{B}$. 

\section{General numerical aspects}
In this section we delineate the general framework for the numerical
solution  of the coupled equations (\ref{eq13}), (\ref{eq14}) and
(\ref{eq15}).  Let us assume certain spatial 
discretizations of the magnetic field in the volume of the dynamo, 
and of the electric potential and the vector potential at the boundary. 
Then,
Eq. (\ref{eq13}) may be rewritten in the form 
\begin{eqnarray}{\label{eq15a}}
B_i=L_{ik}B_k+\lambda M_{ik}B_k+
\lambda P_{in}A_n+N_{il}\varphi_l \; ,
\end{eqnarray}
where the $B_i$ denote the degrees of freedom of  the magnetic field in the
volume of the dynamo,  while  $\varphi_l$  and $A_n$ denote the
degrees of freedom of the electric potential and the vector potential 
at the boundary. Here and in the following we use Einstein's summation
convention, and we reserve the indices $i,k$ for 
magnetic field degrees of freedom in the volume of the
fluid, whereas the indices $l,m,n$ are reserved for the vector potential and 
electric potential degrees of freedom at the boundary of the fluid.

For any given dynamo source, any shape of the dynamo domain, and any concrete
form of the  discretization, the matrices ${\bf{L}}$, ${\bf{M}}$, 
${\bf{N}}$ and ${\bf{P}}$ in Eq. (\ref{eq15a}) 
can easily be derived from Eq. (\ref{eq13}). 
It is worthwhile to note that only ${\bf{L}}$ depends on the dynamo source
(${\bf{u}}$ or $\alpha$), whereas ${\bf{M}}$, 
${\bf{N}}$ and ${\bf{P}}$ depend only on the
geometry of the dynamo domain and the discretization details.

\begin{figure}
\epsfxsize=12cm\epsfbox{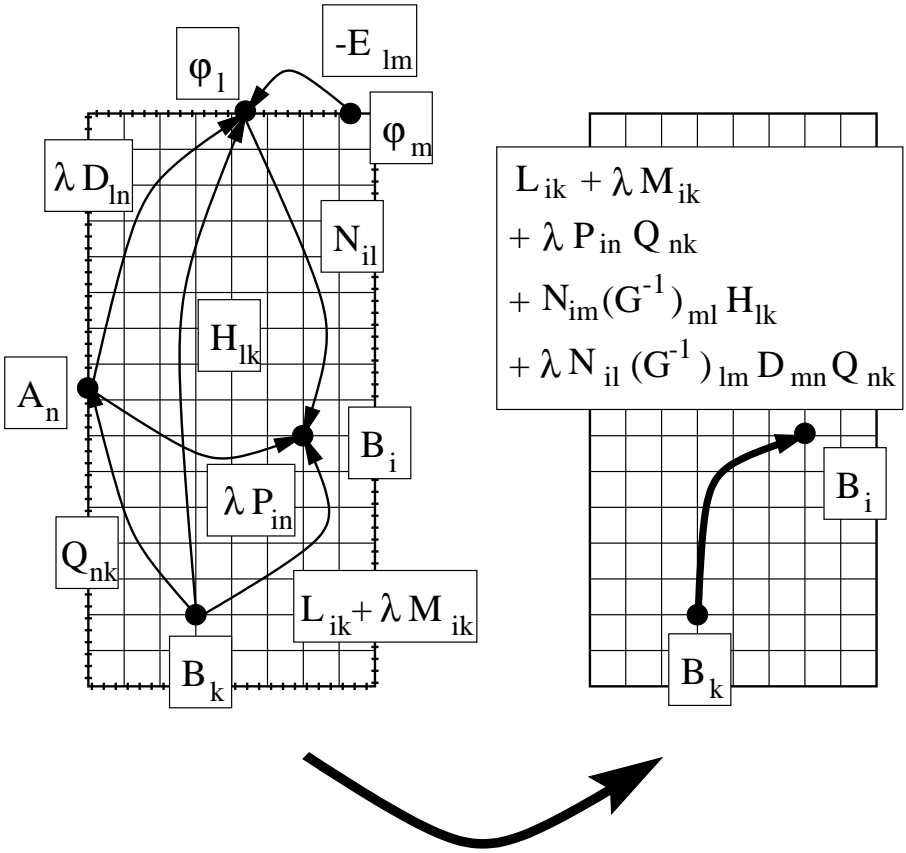}
\caption{The general scheme  for the numerical realization of the
integral equation approach. Left: The connection between the magnetic fields in the
volume  and the vector potential and the electric potential 
on the boundary. Right: Reduction to the pure connection between magnetic fields
in the volume. The boundary on the l.h.s. is indicated to have a finer
discretization then the volume. On the r.h.s. the influence of the boundary has been
incorporated implicitly. For details, cp the formulae in the text.}
\label{fig0} 
\end{figure}

Similarly, 
the discretization of the boundary integral equation (\ref{eq14}) 
(for the case that ${\bf{r}}$ is on the surface $S$) leads to
\begin{eqnarray}{\label{eq15b}}
0.5 \; \varphi_l+E_{lm}\varphi_m=H_{lk}B_k+\lambda D_{ln}A_n,
\end{eqnarray}
or
\begin{eqnarray}{\label{eq15c}}
G_{lm}\varphi_m=H_{lk}B_k+\lambda D_{ln}A_n,
\end{eqnarray}
where $G_{lm}=0.5 \; \delta_{lm}+E_{lm}$.  
Again one should note that only ${\bf{H}}$ depends on the dynamo source, 
whereas ${\bf{G}}$ and ${\bf{D}}$ depend only on the
geometry of the dynamo domain and the discretization details.
The discretization of Eq. (\ref{eq15}) gives
\begin{eqnarray}{\label{eq15d}}
A_{n}=Q_{nk}B_k \; ,
\end{eqnarray}
with $\bf Q$ depending solely on the geometry.
The influence of the magnetic field degrees of freedom on the vector potential
and the electric potential degrees of freedom on the boundary, and vice versa,
is illustrated on the l.h.s. of Fig. \ref{fig0}.

Substituting Eq. (\ref{eq15d}) into Eq. (\ref{eq15c}) yields
\begin{eqnarray}{\label{eq15e}}
\varphi_m=(G^{-1})_{ml}H_{lk}B_k+\lambda (G^{-1})_{ml}D_{ln}Q_{nk}B_k.
\end{eqnarray}
However, for the inversion of the matrix $\bf G$ some care is needed.
Basically, $\bf G$ is a singular matrix, reflecting the fact that the
electric potential is determined only up to a constant. 
Accidently, it may happen 
that this singularity is weakened by inaccuracies due to
the discretization. 
Nevertheless, one should be careful with the inversion. A convenient 
method to deal with the inversion is the application of the 
{\it deflation method} \cite{BARN,SARV}. In the following, we will simply
assume that the  inverse of $\bf G$ has been found in an appropriate manner.

After inserting Eqs. (\ref{eq15d}) and (\ref{eq15e}), Eq.
(\ref{eq15a}) is  transformed to
\begin{eqnarray}{\label{eq15f}}
B_i=L_{ik}B_{k}+\lambda 
M_{ik}B_{k}+\lambda P_{in}Q_{nk}B_k+N_{im}(G^{-1})_{ml}H_{lk}B_k+
\lambda N_{il}(G^{-1})_{lm}D_{mn}Q_{nk}B_k.
\end{eqnarray}
Evidently, the electric potential and the magnetic vector potential
at the boundary served only  as auxiliary quantities in order 
to ensure the 
right boundary conditions. 
From the numerical viewpoint it is important 
to take notice of the following: for the accurate solution of the boundary
integral equation (\ref{eq15e}) for a dynamo in a given domain 
it may be advisable to use a fine discretization 
of the boundary with a large number of grid points. Hence, the corresponding
inversion  of the matrix  $\bf G$ might be numerically expensive. However, for
a given  geometry this inversion is needed only once. Finally, after carrying
out  the matrix multiplications $\bf N \cdot G^{-1} \cdot H$ and 
$\bf N \cdot G^{-1} \cdot D \cdot Q$ in Eq. (\ref{eq15f}) 
one ends up with
a matrix of the order $(NB,NB)$ where $NB$ denotes the total number 
of all magnetic field degrees of freedom (see the r.h.s. of 
Fig. (\ref{fig0})).

Now, Eq. (\ref{eq15f}) can be rewritten in the following form:
\begin{eqnarray}{\label{eq15g}}
(\delta_{ik}-L_{ik}-N_{im}(G^{-1})_{ml}H_{lk})B_k=\lambda (M_{ik}+P_{in}Q_{nk}+
N_{il}(G^{-1})_{lm}D_{mn}Q_{nk})B_k.
\end{eqnarray}
This is a generalized linear matrix
eigenvalue problem in which only the magnetic field components
remain. 
The numerical solution of the arising linear generalized eigenvalue 
equation (\ref{eq15g})
yields the eigenvalues $\lambda$, comprising as the real part 
the growth rate and as the imaginary part 
the frequency  of the dynamo.

\section{Application to time-dependent $\alpha^2$ dynamos - Basics}
In this section, we exemplify the general integral 
equation approach by applying it
to a simple mean-field dynamo model with a spherically symmetric, isotropic
helical turbulence parameter $\alpha$. 
In contrast to the original model with constant $\alpha$ \cite{KRST,KRRA},
whose advantage is the possibility of an analytical treatment,
we allow here $ \alpha$ to  vary with the radial coordinate $r$.
Starting from equations (\ref{eq13}), (\ref{eq14}) and (\ref{eq15}) we will 
derive two coupled radial integral equations which will be used in the next section
for numerical treatment. Note that for the steady case the corresponding
derivation had been given in \cite{STE2}.

Basically, this section is intended as an exercise to show how 
the basic integral equation approach can be reduced 
in a particular  case with high symmetry of the dynamo source. A slightly
simpler derivation of  the two radial integral equations,
which 
starts from the radial differential equations and  
uses a Green's function method, is
given in Appendix A. 
If the reader is only  interested in the final form of the radial integral
equations he can skip the following derivations and refer directly  to 
Eqs. (\ref{eq27}) and (\ref{eq41}).

\subsection{Preliminaries}
As usual in dynamo theory, we split the divergence-free magnetic field
$\bf{B}$  into a poloidal and a toroidal part, denoted by ${\bf{B}}_{P}$ and
${\bf{B}}_{T}$. Since we use the Coulomb gauge, $\nabla \cdot {\bf A}=0$, an
equivalent  decomposition can also be applied to the vector potential, 
${\bf A}={\bf{A}}_{P}+{\bf{A}}_{T}$. 
We represent these fields by the defining scalars $S$, $T$, $S^A$, $T^A$
according to  
\begin{eqnarray}{\label{eq16}}
{\bf{B}}_{P}=\nabla \times \nabla \times\left(\frac{S}{r} \, {\bf{r}} 
\right),
\; \; \; 
{\bf{B}}_{T}=\nabla\times \left(\frac{T}{r} \, {\bf{r}} \right) \; ,
\end{eqnarray}
\begin{eqnarray}{\label{eq16a}}
{\bf{A}}_{P}=\nabla \times \nabla \times\left(\frac{S^A}{r} \, {\bf{r}} 
\right),
\; \; \; 
{\bf{A}}_{T}=\nabla\times \left(\frac{T^A}{r} \, {\bf{r}} \right) \; .
\end{eqnarray}
We introduce spherical coordinates $r,\theta,\phi$ and denote the 
radius vector by $\bf{r}$.
The defining scalars and the electric potential 
are expanded in series of spherical harmonics $Y_{lm}(\theta,\phi)$,
\begin{eqnarray}{\label{eq17}}
S(r,\theta,\phi)&=&\sum_{l,m} s_{lm}(r) Y_{lm}(\theta,\phi)  ,\;\;\;\nonumber\\
T(r,\theta,\phi)&=&\sum_{l,m} t_{lm}(r) Y_{lm}(\theta,\phi)  ,\;\;\;\nonumber\\
S^A(r,\theta,\phi)&=&\sum_{l,m} s^A_{lm}(r) Y_{lm}(\theta,\phi)  ,\;\;\;\\
T^A(r,\theta,\phi)&=&\sum_{l,m} t^A_{lm}(r) Y_{lm}(\theta,\phi)  ,\;\;\;\nonumber\\
\varphi(r,\theta,\phi)&=&\sum_{l,m} \varphi_{lm}(r) Y_{lm}(\theta,\phi) \; .\nonumber
\end{eqnarray}
Later $S^A$, $T^A$, $\varphi$ will be needed only at the boundary $r=R$.
For the spherical harmonics $Y_{lm}(\theta,\phi)$ 
the definition
\begin{eqnarray}{\label{eq18}}
Y_{lm}(\theta,\phi)=\sqrt{\frac{2l+1}{4 \pi}\frac{(l-m)!}{(l+m)!}} \; 
P_{lm}(\cos \theta) e^{im \phi} \
\end{eqnarray}
is employed, with $P_{lm}$ denoting associated Legendre Polynomials. The
summations  in (\ref{eq17}) are over all degrees $l$ and orders $m$ satisfying
$l\ge 0$ and $|m| \le l$;  terms with $l=0$, however, are without interest in
the following.  Since $S$, $T$, $S^A$, $T^A$, and $\varphi$ are real we have
$s_{l-m}=s_{lm}^\ast$ and  analogous relations for $t_{lm}$, $s^A_{lm}$,
$t^A_{lm}$  and $\varphi_{lm}$. The definition (\ref{eq18}) implies
the following orthogonality relation for the $Y_{lm}(\theta,\phi)$:
\begin{eqnarray}{\label{eq19}}
\int_0^{2\pi} d\phi \int_0^{\pi} \sin \theta \, d 
\theta \, Y_{l'm'}^{\ast}(\theta,\phi)
Y_{lm}(\theta,\phi)=\delta_{ll'} \delta_{mm'} \; .
\end{eqnarray}
A useful relation is
\begin{eqnarray}{\label{eq19}}
\Omega \; Y_{lm}=-l(l+1) Y_{lm}\; ,
\end{eqnarray}
where the operator $\Omega$ is defined by
\begin{eqnarray}{\label{eq20}}
\Omega f=\frac{1}{\sin \theta} \, \frac{\partial}{\partial 
\theta} \left(\sin \theta
\; \frac{\partial f}{\partial \theta} \right)+\frac{1}{\sin^2 
\theta} \; 
\frac{\partial^2 f}
{\partial \phi^2} \;.
\end{eqnarray}
From Eqs. (\ref{eq16}) , (\ref{eq16a}), and (\ref{eq17}) we obtain, with the
help of (\ref{eq19}),  the components of 
$\bf{B}$
\begin{eqnarray}\label{eq21b}
B_r(r,\theta,\phi)&=&\sum\limits_{l,m} \frac{l(l+1)}{r^2} s_{lm} (r) 
Y_{lm}(\theta,\phi)   \label{eq21}\nonumber\\
B_{\theta}(r,\theta,\phi)&=&\sum\limits_{l,m} \left(
\frac{t_{lm}(r)}{r \sin \theta} \frac{\partial Y_{lm}(\theta,\phi)}
{\partial \phi}+
\frac{1}{r} \frac{d s_{lm} (r)}{d r}
\frac{\partial Y_{lm}(\theta,\phi)}
{\partial \theta} \right)  \label{eq21d}\\
B_{\phi}(r,\theta,\phi)&=&\sum\limits_{l,m} \left(  
- \frac{t_{lm}(r)}{r} \frac{\partial Y_{lm}(\theta,\phi)}
{\partial \theta}+\frac{1}{r \sin \theta} \frac{d s_{lm} (r)}{d r}
\frac{\partial Y_{lm}(\theta,\phi)}
{\partial \phi} \right) \; , \nonumber
\end{eqnarray}
and of $\bf{A}$
\begin{eqnarray}{\label{eq21cc}}
A_r(r,\theta,\phi)&=&\sum\limits_{l,m} \frac{l(l+1)}{r^2} s^A_{lm} (r) 
Y_{lm}(\theta,\phi)  {\label{eq21aa}} \nonumber \\
A_{\theta}(r,\theta,\phi)&=&\sum\limits_{l,m} \left(
\frac{t^A_{lm}(r)}{r \sin \theta} \frac{\partial Y_{lm}(\theta,\phi)}
{\partial \phi}+
\frac{1}{r} \frac{d s^A_{lm} (r)}{d r}
\frac{\partial Y_{lm}(\theta,\phi)}
{\partial \theta} \right) {\label{eq21bb}} \\
A_{\phi}(r,\theta,\phi)&=&\sum\limits_{l,m} \left(  
- \frac{t^A_{lm}(r)}{r} \frac{\partial Y_{lm}(\theta,\phi)}
{\partial \theta}+\frac{1}{r \sin \theta} \frac{d s^A_{lm} (r)}{d r}
\frac{\partial Y_{lm}(\theta,\phi)}
{\partial \phi} \right)  \; .\nonumber 
\end{eqnarray}

Finally, we recall the expression for the inverse distance
between two points $\bf{r}$ and $\bf{r'}$, 
\begin{eqnarray}{\label{eq22}}
\frac{1}{|{\bf{r}}-{\bf{r'}}|}=4 \pi \sum_{l=0}^{\infty} 
\sum_{m=-l}^{l}
\frac{1}{2l+1} \frac{r_{<}^l}{r_{>}^{l+1}} 
Y_{lm}^{\ast}(\theta',\phi')
Y_{lm}(\theta,\phi) \; ,
\end{eqnarray}
where $r_{>}$ denotes the larger of the values $r$ and $r'$, and
$r_{<}$  the smaller one. 

Equipped with these preliminaries, we will derive 
two coupled integral equations 
for the functions $s_{lm}(r)$ and 
$t_{lm}(r)$. To meet this goal, we will also need expressions 
for $s^A_{lm}(r)$ and $\varphi(r)$  which are necessary, however, only at some
intermediate steps.  

\subsection{Radial integral equation for $s_{lm}(r)$}
Taking the scalar product of both sides  of Eq. (\ref{eq13}) with the unity
vector ${\bf{e}}_r$ we obtain 
\begin{eqnarray}\label{eq23}\
{\bf{B}}({\bf{r}}) \cdot {\bf{e}}_r&=&
\frac{\mu_0 \sigma}{4 \pi} \int_D
\frac{  (\alpha({{r'}}) {\bf{B}}({\bf{r'}})-\lambda {\bf{A}}({\bf{r}}')) \times 
({\bf{r}}-{\bf{r'}})}{|{\bf{r}}-{\bf{r'}}|^3} \cdot {\bf{e}}_r\; dV'
\nonumber\\ 
&&-\frac{\mu_0 \sigma}{4 \pi} \int_S 
\varphi({\bf{s'}}) 
{\bf{n}} ({\bf{s'}}) \times
\frac{{\bf{r}}-{\bf{s'}}}{|{\bf{r}}-{\bf{s'}}|^3}
\cdot {\bf{e}}_r \; dS' \nonumber\\
&=&\frac{\mu_0 \sigma}{4 \pi} \int_D
\frac{\nabla_{r'}  \times  
(\alpha({{r'}}) {\bf{B}}({\bf{r'}})-\lambda{\bf{A}}({\bf{r}}'))}{|{\bf{r}}-{\bf{r'}}|} 
\cdot {\bf{e}}_{r'} \; \frac{r'}{r} \; dV' \; .\\
\nonumber
\end{eqnarray}
In the derivation of the last step in Eq. (\ref{eq23}) we have expressed
 ${\bf{e}}_{r}$ under the integrals by
$({\bf{r}}-{\bf{r'}})/r +(r'/r)  {\bf{e}}_{r'}$, and we have used the fact that
the triple product vanishes 
when $ {\bf{n}} ({\bf{s'}})$ and ${\bf{e}}_{r'}$ coincide for
$\bf{r'}=\bf{s'}$.

By virtue of Eq. (\ref{eq10}), Eq. (\ref{eq23}) becomes
\begin{eqnarray}{\label{eq23a}}
 {\bf{B(r)}}\cdot {\bf{e}}_r=\frac{\mu_0 \sigma}{4 \pi} \int_D
\frac{\nabla_{r'}  \times  
(\alpha({{r'}}) {\bf{B}}({\bf{r'}}))}{|{\bf{r}}-{\bf{r'}}|} 
\cdot {\bf{e}}_{r'} \; \frac{r'}{r} \; dV' \; -\frac{\mu_0 \sigma\lambda}
{4 \pi}\int_D\frac{ {\bf{B}}({\bf{r'}})}{|{\bf{r}}-{\bf{r'}}|} 
\cdot {\bf{e}}_{r'} \; \frac{r'}{r} \; dV' \; .
\end{eqnarray}
Noting that in Eq. (\ref{eq23a}) 
\begin{eqnarray}{\label{eq24}}
\nabla_{r'}  \times  
(\alpha({{r'}}) {\bf{B}}({\bf{r'}}))= -{\bf{B}}({\bf{r'}}) \times 
\nabla_{r'} \alpha({{r'}})+\alpha({{r'}}) \nabla_{r'} \times
{\bf{B}}({\bf{r'}})\; ,
\end{eqnarray}
we see  that the scalar product of the first term
on the right hand side with ${\bf{e}}_{r'}$ vanishes since the gradient
of $\alpha({r'})$ points in $\bf{r'}$-direction, too.
 From Eqs. (\ref{eq19}) and  (\ref{eq21b}) we obtain
\begin{eqnarray}{\label{eq25}}
(\nabla_{r'} \times {\bf{B}}({\bf{r'}}))\cdot
{\bf{e}}_{r'}&=&\sum\limits_{l',m'} 
\frac{l'(l'+1)}{{r'}^2} t_{l'm'} (r') 
Y_{l'm'}(\theta',\phi') \; .
\end{eqnarray}
Taking Eqs. (\ref{eq23}), (\ref{eq24}) and (\ref{eq25}) together we find
\begin{eqnarray}
\sum_{l,m} \frac{l(l+1)}{r^2} 
s_{lm}(r) Y_{lm}(\theta,\phi)&=& 
\frac{\mu_0 \sigma}{4 \pi r} 
\int_D \alpha(r')
\sum_{l'm'}\frac{l'(l'+1)}{{r'}^2} t_{l'm'} (r') 
Y_{l'm'}(\theta',\phi') \frac{r'}{|\bf{r}-\bf{r'}|} dV'{\label{eq26}}
\nonumber \\ 
 &&-\frac{\mu_0\sigma\lambda}{4\pi
r}\int_D\sum_{l'm'}\frac{l'(l'+1)}{{r'}^2}
s_{l'm'}(r')Y_{l'm'}(\theta',\phi')\frac{r'}{|{\bf{r}}-{\bf{r}}'|}dV'
\; .
\end{eqnarray} 
After expressing the inverse distance according to Eq.
(\ref{eq22}), integrating 
on the
right-hand side of Eq. (\ref{eq26})
over the primed angles, multiplying then 
both sides of  Eq. (\ref{eq26}) with $Y_{lm}^{\ast}(\theta,\phi)$ and
integrating over the non-primed angles we obtain the first 
integral equation 
of our problem in the form
\begin{eqnarray}
s_{lm}(r)&=&\frac{\mu_0 \sigma }{2l+1} [
\int_0^r  \frac{{r'}^{l+1}}{r^{l}} \, \alpha(r') \, t_{lm}(r') 
\, dr'+
\int_r^R  \frac{r^{l+1}}{{r'}^{l}} \, \alpha(r') \, t_{lm}(r') \, 
dr' \nonumber\\
&&-\lambda\int_0^r\frac{{r'}^{l+1}}{r^l}s_{lm}(r')dr'-\lambda\int_r^R
\frac{r^{l+1}}{{r'}^l}s_{lm}(r')dr'] \; .{\label{eq27}}\
\end{eqnarray}
\subsection{The electric potential at the boundary}
For the determination of the electric potential at the boundary it is convenient
to start
from Eq. (\ref{eq14}) for points $\bf{r}$ outside $D$. As for the last
boundary  integral in Eq. (\ref{eq14}) we have 
\begin{eqnarray}{\label{eq28}}
\frac{1}{4 \pi} \int\limits_S \varphi({\bf{s'}})
{\bf{n}}({\bf{s'}}) 
\cdot \frac{{\bf{r}}-{\bf{s'}}}{{|{\bf{r}}-{\bf{s'}}|}^3} \;  dS' 
=\frac{1}{4 \pi} \int\limits_S \varphi({\bf{s'}})
\frac{\partial}{\partial s'}
\frac{1}{{|{\bf{r}}-{\bf{s'}}|}} \; dS' \nonumber
\end{eqnarray}
\begin{eqnarray}{\label{eq29}}
=\int\limits_S \sum_{lm} \varphi_{lm}(R) Y_{lm}(\theta',\phi')
\sum_{l'm'} \frac{1}{2l'+1} \frac{\partial}{\partial s'}
\frac{{s'}^{l'}}{r^{l'+1}}
Y_{l'm'}^{\ast}(\theta',\phi') Y_{l'm'}(\theta,\phi) \; dS' 
\end{eqnarray}
and thus
\begin{eqnarray}{\label{eq30}}
\lim_{\bf{r}\to\bf{s}} \frac{1}{4 \pi} \int\limits_S 
\varphi({\bf{s'}})
{\bf{n}}({\bf{s'}}) 
\cdot \frac{{\bf{r}}-{\bf{s'}}}{{|{\bf{r}}-{\bf{s'}}|}^3} \;  dS'=
\sum_{lm} \frac{l}{2l+1} \varphi_{lm}(R) Y_{lm}(\theta,\phi) \;.
\end{eqnarray}
For the evaluation of the volume integrals in Eq. (\ref{eq14}) we can make
the second one vanish by means of the Coulomb gauge $\nabla \cdot {\bf A}$.
For the first one, we use the alternative formulation Eq. (\ref{eq14aa}).
Taking 
 $B_r$ from Eq. (\ref{eq21}), we find  
\begin{eqnarray}{\label{eq32}}
\int\limits_D  
\frac{\nabla_{r'} \cdot (\alpha(r') 
{\bf{B}}({\bf{r'}}))}{|{\bf{r}}-{\bf{r'}}|} 
dV' 
&=&\int\limits_D
\frac{d \alpha(r')}{d r'} \sum_{l'm'} \frac{l'(l'+1)}{{r'}^2}
s_{l'm'}(r') Y_{l'm'}(\theta',\phi') \frac{1}{|{\bf{r}}-{\bf{r'}}|} 
dV'
\end{eqnarray}
and thus
\begin{eqnarray}{\label{eq33}}
\lim_{\bf{r}\to\bf{s}} \frac{1}{4 \pi} \int\limits_D  
\frac{\nabla_{r'} \cdot (\alpha(r') 
{\bf{B}}({\bf{r'}}))}{|{\bf{r}}-{\bf{r'}}|} 
dV'&=& \sum_{lm} \frac{l(l+1)}{2l+1}  Y_{lm}(\theta,\phi)  
\int_0^R  \frac{{r'}^l}{R^{l+1}} \,
\frac{d \alpha(r')}{d r'} 
s_{lm}(r') \, dr' \; .
\end{eqnarray}
Analogously, we obtain for the remaining boundary integrals in Eq.
(\ref{eq14})   \begin{eqnarray}{\label{eq34}}
\lim_{\bf{r}\to\bf{s}}\frac{1}{4 \pi} \int\limits_S
{\bf{n}}({\bf{s'}}) \cdot
\frac{\alpha({{s'}}) {\bf{B}}({\bf{s'}})-
\lambda{\bf{A}}({\bf{s}}')}{|{\bf{r}}-{\bf{s'}}|} 
\; dS'&=&
\sum_{lm} \frac{l(l+1)}{2l+1} \, \frac{1}{R} \, (\alpha(R)  \, 
s_{lm}(R) \,-\lambda\, s^A_{lm}(R))\nonumber \\
&& Y_{lm}(\theta,\phi)\; .
\end{eqnarray}
Evaluating now Eq. (\ref{eq14}) for ${\bf{r}} \rightarrow {\bf{s}}$ with the
help  of Eqs. (\ref{eq30}), (\ref{eq33})
and (\ref{eq34}) we find
\begin{eqnarray}{\label{eq35}}
\varphi_{lm}(R)=-(l+1) \int_0^R \frac{{r'}^l}{R^{l+1}} \,
\frac{d \alpha(r')}{d r'} \, s_{lm}(r') \, dr'
+\frac{l+1}{R} \,( \alpha(R) \, s_{lm}(R)-\lambda\, s^A_{lm}(R)) \; .
\end{eqnarray}
This expression for $\varphi(R)$ will later be needed in the integral equation
for $t_{lm}(r)$.

\subsection{The vector potential at the boundary}
From Eq. (\ref{eq15}) and the fact that the curl of the magnetic
field  vanishes outside of $D$, we obtain
\begin{eqnarray}{\label{eqp1}}
\int_D\frac{\nabla_{r'}\times {\bf{B}}({\bf{r'}})}{|{\bf{r}}-{\bf{r'}}|}dV'
=\int_{D+D'}\frac{\nabla_{r'}\times {\bf{B}}({\bf{r'}})}{|{\bf{r}}-{\bf{r'}}|}dV',
\end{eqnarray}
where $D'$ denotes the outside region of $D$. Note that
\begin{eqnarray}{\label{eqp1a0}}
\nabla_{r'}\times\frac{{\bf{B}}({\bf{r'}})}{|{\bf{r}}-{\bf{r'}}|}=\nabla_{r'}
\frac{1}{|{\bf{r}}-{\bf{r'}}|}\times 
{\bf{B}}({\bf{r'}})+\frac{1}{|{\bf{r}}-{\bf{r'}}|}
\nabla_{r'}\times {\bf{B}}({\bf{r'}}).
\end{eqnarray}
The application of this equation on the right side of the Eq. (\ref{eqp1}) 
leads  to
\begin{eqnarray}{\label{eqp1a}}
\int_{D}\frac{\nabla_{r'}\times {\bf{B}}({\bf{r'}})}{|{\bf{r}}-{\bf{r'}}|}dV'=
\int_{D+D'}\nabla_{r'}\times \frac{{\bf{B}}({\bf{r'}})}{|{\bf{r}}-{\bf{r'}}|}dV'-
\int_{D+D'}\nabla_{r'}\frac{1}{|{\bf{r}}-{\bf{r'}}|}\times {\bf{B}}({\bf{r'}})dV'.
\end{eqnarray}
Applying Gauss' theorem, we have
\begin{eqnarray}{\label{eqp1b}}
\int_D\frac{\nabla_{r'}\times {\bf{B}}({\bf{r'}})}{|{\bf{r}}-{\bf{r'}}|}dV'=
\int_{S_{\infty}}
{\bf{n}}({\bf{s'}})\times\frac{{\bf{B}}({\bf{r'}})}{|{\bf{r}}-{\bf{r'}}|}
dS'-\int_{D+D'}\frac{{\bf{r}}-{\bf{r'}}}{|{\bf{r}}-{\bf{r'}}|^3}\times
{\bf{B}} ({\bf{r'}})dV'. \end{eqnarray}
Equation (\ref{eq3}) allows to conclude that the surficial integration 
on the r.h.s. of this equation vanishes. Therefore, after taking the scalar 
product of both sides of the equation (\ref{eqp1b}) with the unit vector 
${\bf{e}}_r$, we obtain
\begin{eqnarray}{\label{eqp2}}
\int_D\frac{\nabla_{r'}\times {\bf{B}}({\bf{r'}})}
{|{\bf{r}}-{\bf{r'}}|}\cdot {\bf{e}}_rdV'=-\int_{D+D'}\nabla_{r'}
\frac{1}{|{\bf{r}}-{\bf{r'}}|}\times {\bf{B}}({\bf{r'}})\cdot {\bf{e}}_{r'}
\frac{r'}{r}dV'.
\end{eqnarray}
In the derivation of this equation, the relation ${\bf{e}}_r=({\bf{r}}-{\bf{r'}})/r+
(r'/r){\bf{e}}_{r'}$ has been used again.
Applying Eq. (\ref{eqp1a0})  again we have
\begin{eqnarray}{\label{eqp3a}}
\int_D\frac{\nabla_{r'}\times {\bf{B}}({\bf{r'}})}{|{\bf{r}}-{\bf{r'}}|}\cdot 
{\bf{e}}_r dV'&=&\int_{D}\frac{\nabla_{r'}\times 
{\bf{B}}({\bf{r'}})}{|{\bf{r}}-{\bf{r'}}|}\cdot {\bf{e}}_{r'}\frac{r'}{r}dV'
-\int_{D+D'}\hspace{-1em}\nabla_{r'}\times\frac{{\bf{B}}({\bf{r'}})}{|{\bf{r}}
-{\bf{r'}}|}\cdot{\bf{e}}_{r'}\frac{r'}{r}dV'
\end{eqnarray}
The second term on the r.h.s. of Eq. (\ref{eqp3a}) can be shown to vanish, so we
conclude 
that
\begin{eqnarray}{\label{eqp2}}
{\bf{A}}({\bf{r}})\cdot {\bf{e}}_r=\frac{1}{4\pi}\int_{D}\frac{\nabla_{r'}\times 
{\bf{B}}({\bf{r'}})}{|{\bf{r}}-{\bf{r'}}|}\cdot {\bf{e}}_{r'}\frac{r'}{r}dV'.
\end{eqnarray}
Combining Eqs.  (\ref{eq21aa}), (\ref{eq22}) and (\ref{eq25}), we obtain
\begin{eqnarray}{\label{eqp3}}
s^A_{lm}(r)=\frac{1}{2l+1}(\int_0^r\frac{{r'}^{l+1}}{r^l}t_{lm}(r')dr'+
\int_r^R\frac{r^{l+1}}{{r'}^l}t_{lm}(r')dr').
\end{eqnarray}
Particularly, when $r=R$ in the above equation, we have
\begin{eqnarray}{\label{eqp4}}
s^A_{lm}(R)=\frac{1}{2l+1}\int_0^R \frac{{r'}^{l+1}}{R^l}t_{lm}(r')dr'.
\end{eqnarray}

As we will see, there is no need to calculate the corresponding expression for
$t^A_{lm}(r)$.

\subsection{Radial integral equation for $t_{lm}(r)$}
We take now the $curl$ of both sides of Eq. (\ref{eq12a}) thus obtaining
\begin{eqnarray}{\label{eq36}}
\nabla_{r} \times {\bf{B}}({\bf{r}})&=&\frac{\mu_0 
\sigma}{4 \pi} [
\nabla_{r} \times \nabla_{r} \times 
\int_D
\frac{\alpha({{r'}}) {\bf{B}}({\bf{r'}})-\lambda {\bf{A}}({\bf{r'}})}
{|{\bf{r}}-{\bf{r'}}|} 
dV' \nonumber
\\
&&-\nabla_{r} \times  \int_S 
\varphi({\bf{s'}}) {\bf{n}} ({\bf{s'}}) \times
\frac{{\bf{r}}-{\bf{s'}}}{|{\bf{r}}-{\bf{s'}}|^3} \; dS'  \;] .
\end{eqnarray}

Considering first the case $r \le R$ we take  on both sides 
of Eq. (\ref{eq36}) the scalar product with ${\bf{e}}_r$.
We note that
\begin{eqnarray}{\label{eq37}}
&&{\bf{e}}_r \cdot \left( \nabla_{r} \times \nabla_{r} \times \int_D
\frac{\alpha({{r'}}) {\bf{B}}({\bf{r'}})-\lambda {\bf{A}}({\bf{r'}})}
{|{\bf{r}}-{\bf{r'}}|} dV' \right)\nonumber\\
&&={\bf{e}}_r \cdot \left(
(\nabla_{r} \nabla_{r} \cdot - \Delta_{r})
\int_D
\frac{\alpha({{r'}}) {\bf{B}}({\bf{r'}})-\lambda {\bf{A}}({\bf{r'}})}
{|\bf{r}-\bf{r'}|} dV' \right)\nonumber\\ 
&&=\frac{\partial}{\partial r} \int_D
\frac{\nabla_{r'} \cdot (\alpha({{r'}}) {\bf{B}}({\bf{r'}}))}
{|{\bf{r}}-{\bf{r'}}|} dV'-\frac{\partial}{\partial r} 
\int\limits_S
{\bf{n}}({\bf{s'}}) \cdot
\frac{\alpha({{s'}}) {\bf{B}}({\bf{s'}})-\lambda {\bf{A}}({\bf{s'}})}{|{\bf{r}}-
{\bf{s'}}|} 
\; dS'\nonumber\\
&&+4 \pi (
\alpha({{r}}) {{B}}_r(r)-\lambda A_r(r))  \; ,
\end{eqnarray}
where we have used the identity $\Delta_{r} |{\bf{r}}-{\bf{r'}}|^{-1}=
-4 \pi \delta({\bf{r}}-{\bf{r'}})$ and the Coulomb gauge 
$\nabla\cdot{\bf{A}}=0$.  The two integrals on the third line of
Eq. (\ref{eq37}) were already treated in subsection C. 
Concerning the boundary integral in  Eq. (\ref{eq36}) over the 
electric potential we note that
\begin{eqnarray}{\label{eq38}}
{\bf{e}}_r \cdot \left(\nabla_{r} \times  
\left( {\bf{n}}({\bf{s'}}) \times
\frac{{\bf{r}}-{\bf{s'}}}{|{\bf{r}}-{\bf{s'}}|^{3}} \right) 
\right)=
-\frac{\partial^2}{\partial r \partial s'} 
\frac{1}{|{\bf{r}}-{\bf{s'}}|}  \; .
\end{eqnarray}
Putting everything together
we obtain
\begin{eqnarray}{\label{eq39}}
(\nabla_r \times {\bf{B}}({\bf{r}})) \cdot {\bf{e}}_r&=&
\mu_0 \sigma (\alpha(r) B_r({\bf{r}})-\lambda A_r(r))+
\frac{\mu_0 \sigma}{4 \pi}  \bigg[  \frac{\partial}
{\partial r} \int_D  
{\frac{d \alpha(r')}{dr'} B_r({\bf{r'}})} \frac{1}{|{\bf{r}}-{\bf{r'}}|} dV'
\nonumber
\\ &&-
\frac{\partial}{\partial r}
\int\limits_S
\frac{\alpha({{s'}}) {{B}}_r({\bf{s'}})-\lambda A_r({\bf{s'}})}{|{\bf{r}}-{\bf{s'}}|} 
\; 
dS' \nonumber +\int_S \varphi({\bf{s'}})
\frac{\partial^2}{\partial r \partial s'} \frac{1}{|{\bf{r}}-{\bf{s'}}|}
dS' \; \bigg].
\end{eqnarray}
Representing now the left-hand side according to Eq. (\ref{eq25}), applying
Eq. (\ref{eq22})  and integrating both sides over 
the angles we
obtain
\begin{eqnarray}
t_{lm}(r)&=&\mu_0\sigma[\alpha(r)s_{lm}(r)-\lambda
s^A_{lm}(r)-\frac{l+1}{2l+1}
\int_0^r\frac{d\alpha(r')}{dr'}s_{lm}(r')\frac{{r'}^l}{r^l}dr'{\label{eq40}}\nonumber\\
&&+\frac{l}{2l+1}\int_r^R\frac{d\alpha(r')}{dr'}s_{lm}(r')
\frac{r^{l+1}}{{r'}^{l+1}}dr'
-\frac{l}{2l+1}\frac{r^{l+1}}{R^{l+1}}(\alpha(R)s_{lm}(R)\nonumber\\
&&-\lambda\; s^A_{lm}(R))-\frac{1}{2l+1}\varphi_{lm}(R)\frac{r^{l+1}}{R^l}] \;.  
\end{eqnarray}
Substituting Eqs. (\ref{eq35}) and (\ref{eqp4}) 
into Eq. (\ref{eq40}) leads to
\begin{eqnarray}
t_{lm}(r)&=&\mu_0\sigma[\alpha(r)s_{lm}(r)-\frac{l+1}{2l+1}\int_0^r\frac{d\alpha(r')}
{dr'}s_{lm}(r')\frac{{r'}^l}{r^l}dr'\label{eq41} \nonumber\\
&&+\frac{l}{2l+1}\int_r^R\frac{d\alpha(r')}{dr'}s_{lm}(r')\frac{r^{l+1}}
{{r'}^{l+1}}dr'+
\frac{l+1}{2l+1}\frac{r^{l+1}}{R^{2l+1}}\int_0^R{r'}^l
\frac{d\alpha(r')}{dr'}s_{lm}(r')dr'\nonumber\\
&&+\frac{\lambda }{2l+1}\frac{r^{l+1}}{R^{2l+1}}\int_0^R{r'}^{l+1}t_{lm}(r')dr'-
\frac{\lambda}{2l+1}\int_0^r\frac{{r'}^{l+1}}{r^l}t_{lm}(r')dr'\nonumber\\
&&-\frac{\lambda}{2l+1}\int_r^R\frac{r^{l+1}}{{r'}^l}t_{lm}(r')dr'-
\frac{r^{l+1}}{R^{l+1}}
\alpha(R)s_{lm}(R)] \; .
\end{eqnarray}
This is the second radial integral equation for our problem. 
Therefore, the  
spherically symmetric $\alpha^2$ dynamo model is reduced to the two
coupled integral  equations (\ref{eq27}) and (\ref{eq41}).  

\subsection{Connection with the differential equation approach}
Notwithstanding the fact that the differential and the integral equation
approach are equivalent in a general sense it might be instructive
to show this equivalence for our special problem. Differentiating
equations (\ref{eq27}) and (\ref{eq41}) two times with respect to the radial component
the following relations can be obtained:
\begin{eqnarray}{\label{eq43}}
\lambda s_{lm}=\frac{1}{\mu_0\sigma}[\frac{d^2s_{lm}}{dr^2}-\frac{l(l+1)}{r^2}s_{lm}]+
\alpha(r)t_{lm},
\end{eqnarray}
\begin{eqnarray}{\label{eq44}}
\lambda t_{lm}=\frac{1}{\mu_0\sigma}[\frac{d^2t_{lm}}{dr^2}-\frac{l(l+1)}{r^2}t_{lm}]-
\frac{d}{dr}(\alpha(r)\frac{ds_{lm}}{dr})+\frac{l(l+1)}{r^2}\alpha(r)s_{lm},
\end{eqnarray}
\begin{eqnarray}{\label{eq45}}
t_{lm}(R)=R\frac{ds_{lm}(r)}{dr}|_{r=R}+l s_{lm}(R)=0.
\end{eqnarray}
As expected, these are  
the differential equations for the considered problem of
radially varying $\alpha$ for the time-dependent case \cite{RAE86,STGE3}.
Note again that from the radial differential equation we can also 
derive the radial integral equations by means of a Green's function technique
(see Appendix A).

\section{Application to time-dependent $\alpha^2$ dynamos - Numerics}

\subsection{Discretization}{\label{DISCRETIZATION}}
In the previous section, and in Appendix A, we have derived the radial
integral  equations (\ref{eq27}) and (\ref{eq41}) governing the time-dependent
dynamo problem. In this subsection, we develop a numerical method to solve 
this integral equations system. 

Let us introduce the following definitions:
\begin{eqnarray}{\label{eq47a}}
x=r/R, C {\alpha}(x)=R^2\mu_0\sigma \alpha(x), 
\tilde{\lambda}_l=R^2\mu_0\sigma \lambda_l,
\end{eqnarray}
where $C$ is the magnitude of the function $R^2\mu_0\sigma \alpha$. In addition, we 
introduce the notations
\begin{eqnarray}{\label{eq48}}
G_s(x,x_0)=\left\{\begin{array}{l}
-\frac{1}{2l+1}x_0^{-l}x^{l+1},\, x\leq x_0,\\
-\frac{1}{2l+1}x_0^{l+1}x^{-l}, x\geq x_0,
\end{array}\right.
\end{eqnarray}
\begin{eqnarray}{\label{eq49}}
G_t(x,x_0)=\left\{\begin{array}{l}
\frac{1}{2l+1}(x_0^{l+1}-x_0^{-l})x^{l+1},\, x\leq x_0, \\
\frac{1}{2l+1}(x^{l+1}-x^{-l})x_0^{l+1}, x\geq x_0,\\
\end{array}\right.
\end{eqnarray}
\begin{eqnarray}{\label{eq50}}
\overline{G}_t(x,x_0)=\left\{\begin{array}{l}
\frac{l+1}{2l+1}x_0^lx^{l+1}+\frac{l}{2l+1}x_0^{-l-1}x^{l+1},x\leq x_0,\\
\frac{l+1}{2l+1}x_0^lx^{l+1}+\frac{l}{2l+1}x^{-l}x_0^l, x\geq x_0.\\
\end{array}\right.
\end{eqnarray}

Then, Eqs. (\ref{eq27}) and (\ref{eq41}) obtain the form
\begin{eqnarray}{\label{eq51}}
s_{lm}(x)=-C\int_0^1 G_s(x,x_0){\alpha}(x_0) t_{lm}(x_0)dx_0+
\tilde{\lambda}_l\int_0^1G_s(x,x_0)s_{lm}(x_0)dx_0,
\end{eqnarray}
\begin{eqnarray}{\label{eq52}}
t_{lm}(x)&=&C {\alpha}(x) s_{lm}(x)-Cx^{l+1}
{\alpha}(1.0)s_{lm}(1.0)+C\int_0^1
\frac{d {\alpha}(x_0)}{dx_0}s_{lm}(x_0)\overline{G}_t(x,x_0)dx_0 \nonumber\\
&&+\tilde{\lambda}_l\int_0^1G_t(x,x_0)t_{lm}(x_0)dx_0-C\int_0^xx^{-l}x_0^l
\frac{d {\alpha}(x_0)}{dx_0}s_{lm}(x_0)dx_0.
\end{eqnarray}
Choosing N equidistant grid points $x_i=i\triangle x$ with $\triangle x=1/N$ 
and approximating the integrals by the extended trapezoidal rule according to
\begin{eqnarray}{\label{eq53}}
\int_0^1f(x)dx\approx\sum_{i=1}^N\frac{1}{2}(f(x_{i-1})+f(x_i))\triangle x,
\end{eqnarray}
we obtain
\begin{eqnarray}{\label{eq54}}
\tilde{\lambda}_l\sum_{j=1}^Ns_{lm}(x_j)G_s(x_i,x_j)\triangle x
c_j=s_{lm}(x_i)+
C\sum_{j=1}^{N-1}{\alpha}(x_j)t_{lm}(x_j)G_s(x_i,x_j)c_j\triangle x,
\end{eqnarray} 
\begin{eqnarray}{\label{eq55}}
\tilde{\lambda}_l\sum_{j=1}^{N-1} G_t(x_i,x_j)t_{lm}(x_j)\triangle x
c_j&=&t_{lm}(x_i)-
C{\alpha}(x_i)s_{lm}(x_i)+Cx_i^{l+1}{\alpha}(1.0)s_{lm}(1.0) \nonumber\\
&&-C\sum_{j=1}^N\frac{d{\alpha}(x_j)}{dx}s_{lm}(x_j)\overline{G}_t(x_i,x_j)
\triangle xc_j\nonumber\\
&&+C\sum_{j=1}^ix_i^{-l}x_j^l\frac{d{\alpha}(x_j)}{dx}s_{lm}(x_j) \triangle x
c_j,\end{eqnarray}
where $c_N=0.5$, and  $c_i=1.0$ for $i=1, 2, \cdots, N-1$.
Equations (\ref{eq54}) and (\ref{eq55}) can be written in the matrix form:
\begin{eqnarray}{\label{eq56}}
\tilde{\lambda}_l{\bf{V}}{\bf{X}}={\bf{W}}{\bf{X}},
\end{eqnarray}
with
\begin{eqnarray}{\label{eq57}}
{\bf{X}}&=&(s_{lm}(x_1),s_{lm}(x_2),\cdots,s_{lm}(x_N),t_{lm}(x_1), t_{lm}(x_2), 
\cdots, t_{lm}(x_{N-1}))^T,\nonumber\\
V_{i,j}&=&G_s(x_i,x_j)\triangle x c_j,(i,j=1, 2, \cdots, N)\nonumber\\ 
V_{i+N,j}&=&0, (i=1, 2, \cdots, N-1, j=1, 2, \cdots, N),\nonumber\\
V_{i,j+N}&=&0, (i=1, 2, \cdots, N, j=1, 2, \cdots, N-1)\nonumber\\ 
V_{i+N,j+N}&=&G_t(x_i,x_j)c_j\triangle x, (i=1, 2, \cdots, N-1, j=1, 2, \cdots, N-1),
\nonumber\\
W_{i,j}&=&\delta(i,j), (i,j=1, 2, \cdots, N)\nonumber\\
W_{i,j+N}&=&G_s(x_i,x_j)\triangle xc_j{\alpha}(x_j), (i=1, 2, 
\cdots, N, j=1, 2, \cdots, N-1)\nonumber\\
W_{i+N, j+N}&=&\delta (i,j), (i, j=1, 2, \cdots, N-1),\nonumber\\
W_{i+N,j}&=&-C\frac{d{\alpha}(x_j)}{dx}\overline{G}_t(x_i,x_j)
\triangle x c_j-C{\alpha}(x_i)\delta(i,j)+Cx_i^{l+1}
{\alpha}(1.0)\delta(j,N)+\nonumber\\
&&Cx_i^{-l}x_j^l\frac{d {\alpha}(x_j)}{dx}\triangle x c_jH(i-j), 
(i=1, 2, \cdots, N-1, j=1, 2, \cdots, N),
\end{eqnarray}
where the functions $\delta(i,j)$ and $H(i-j)$ are defined as
\[
\delta(i,j)=\left\{\begin{array}{l}
0, i\not= j,\\
1, i=j,\\
\end{array}\right.
\]
and 
\[
H(i-j)=\left\{\begin{array}{l}
1, i\geq j,\\
0, i<j.\\
\end{array}\right.
\]
Note that Eq. (\ref{eq56}) is a linear generalized eigenvalue problem. By
multiplying
both sides of Eq. (\ref{eq56}) by the  inverse of the matrix $\bf{V}$, we can
convert it to the following  standard eigenvalue problem:
\begin{eqnarray}{\label{eq58}}
{\bf{V}}^{-1}{\bf{W}}{\bf{X}}=\tilde{\lambda}_l {\bf{X}}.
\end{eqnarray}
This eigenvalue problem can be solved by standard numerical routines. 
First, the matrix ${\bf{V}}^{-1} \bf{W}$ is reduced to the Hessenberg form, 
then the QR algorithm can be employed to obtain the eigenvalue
$\tilde{\lambda}_l$.

\subsection{Numerical results}
In this subsection, we illustrate the numerical performance of 
the integral equation 
approach formulated in this paper by 
a few examples for the functions $\alpha(x)$.

\begin{table}
\caption{Comparison of the calculated growth rates and the analytic ones 
for the free decay case $\alpha(x)=0$. 
The degree of the spherical harmonics is $l=1$. $n=1...4$ correspond to
to modes with increasing radial wavenumber $n$. 
The last row shows the
analytic results. The other  rows express the numerical results obtained by the
integral equation  approach for different grid numbers $N$.}
\label{t1}
\begin{center}
\begin{tabular}{ccccc}\hline
N&n=1&n=2&n=3&n=4\\
\hline
8&-9.79494&-37.71179&-79.61435&-129.36969\\
16&-9.85079&-39.02608&-86.41265&-150.20904\\
32&-9.86485&-39.36475&-88.21583&-155.94894\\
64&-9.86844&-39.44979&-88.67356&-157.42004\\
128&-9.86933&-39.47085&-88.78596&-157.78870\\
\hline
analytic&-9.86960&-39.47842&-88.82644&-157.91367\\
\hline
\end{tabular}
\end{center}
\end{table}

\begin{table}
\caption{Same as Table \ref{t1}, but for $l=2$.}
\label{t2}
\begin{center}
\begin{tabular}{ccccc}\hline
N&n=1&n=2&n=3&n=4\\
\hline
8&-19.87668&-55.87647&-103.28617&-155.45810\\
16&-20.11100&-58.68925&-114.71107&-186.07856\\
32&-20.17073&-59.42935&-117.83333&-194.82947\\
64&-20.18561&-59.61664&-118.63189&-197.09552\\
128&-20.18911&-59.66311&-118.83366&-197.66728\\
\hline
analytic&-20.19073&-59.67951&-118.89986&-197.85780\\
\hline      
\end{tabular}
\end{center}
\end{table}

\begin{figure}
\epsfxsize=12cm\epsfbox{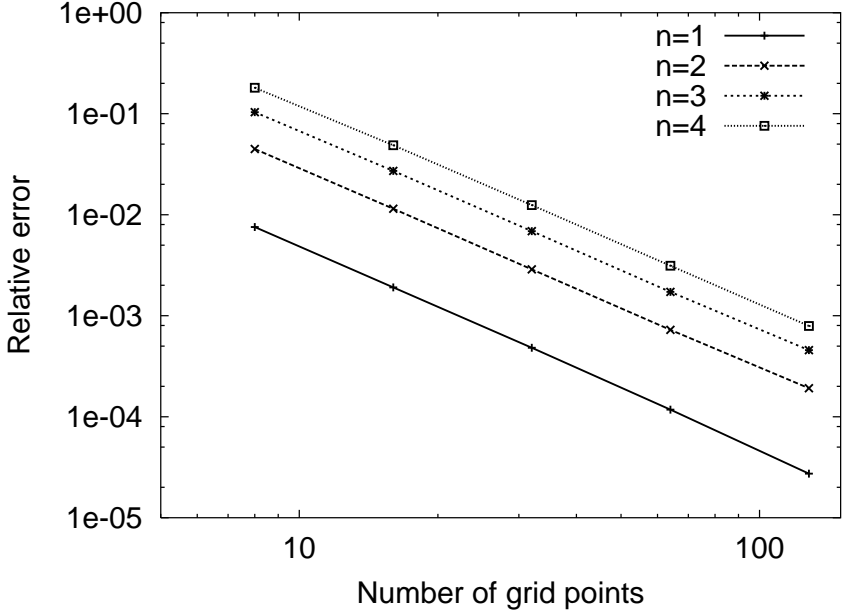}
\caption{Relative error of the numerically determined eigenvalues
for the
free decay case,  $C=0$, $l=1$, $n=1...4$.
The convergence behaves like $N^{-2}$.}
\label{fig1} 
\end{figure}

\begin{figure}
\epsfxsize=12cm\epsfbox{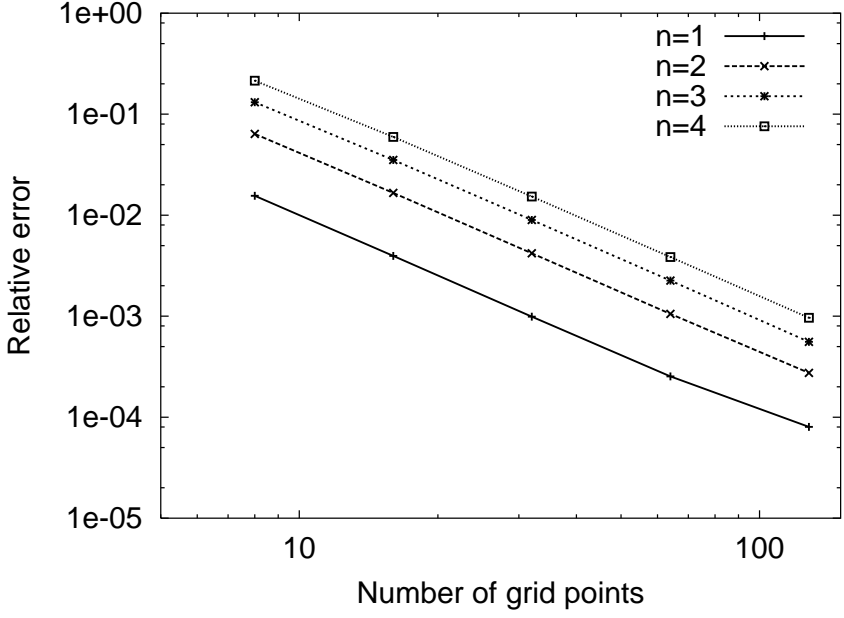}
\caption{Same as Fig. \ref{fig1}, but for $l=2$.}
\label{fig2}
\end{figure}

Let us start with the case $\alpha(x)=0$, which corresponds to a pure 
field decay within a conducting sphere. For this case, the eigenvalues
$\tilde{\lambda}_l$ are known  from quasi-analytic calculations
\cite{KRRA}. In  Table (\ref{t1}) (for $l=1$)  and  Table (\ref{t2})
(for $l=2$) we have
listed the numerical results of the integral equation solver for the 
eigenmodes
with increasing radial wavenumbers  $n=1...4$, together with the
analytical results. From these tables it becomes 
obvious that even for a few grid numbers robust results can be obtained. 
In Figs. \ref{fig1} and \ref{fig2} we have plotted the 
relative errors of the results in dependence on 
the used number of grid points $N$. As it is typical for integral equations of
that kind \cite{XUSG}, the relative error decreases like $N^{-2}$.

\begin{figure}
\epsfxsize=12cm\epsfbox{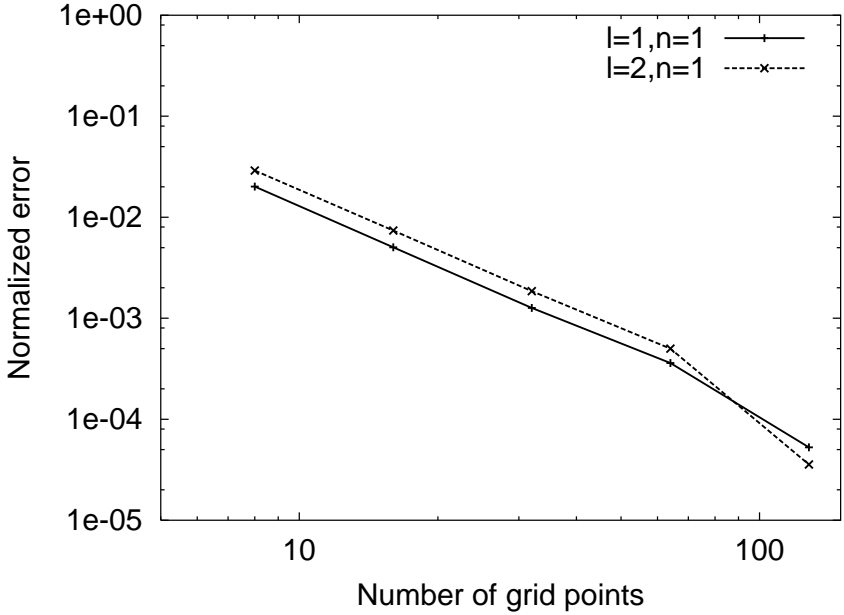}
\caption{Convergence of the first eigenvalue ($n=1$) to zero in 
the cases: $C=4.4934095, \; l=1$ and $C=5.7634593, \; l=2$.
The normalization of the error is done with respect
to the growth rates at $C=0$ which are 
$\lambda_1=-9.8696$ and $\lambda_1=-20.1907$.}
\label{fig3}
\end{figure}

Another quasi-analytic result exists for the steady case \cite{KRRA}: for
$l=1$ and $C=4.4934095$  or $l=2$ and $C=5.7634593$ we know that the first
eigenvalues
have to be  zero.  Table (\ref{t3}) shows the results of the integral equation
solver, again for $l=1$ and $l=2$, but only  for $n=1$.  The convergence of
the results, which is again $\sim N^{-2}$, is depicted  in  Fig. \ref{fig3}.

\newpage

\begin{table}
\caption{Convergence of the first eigenvalue ($n=1$) to zero in 
the cases: $=4.4934095, \; l=1$ and $=5.7634593, \; l=2$.
The normalization of the errors is done with
the growth rates at $C=0$, which are $\lambda_1(C=0)=-9.8696$ and
$\lambda_2(C=0)=-20.1907$}
\label{t3}
\begin{center}
\begin{tabular}{cccccc}
\hline
l&N=8&N=16&N=32&N=64&N=128\\
1&0.19836&0.04971&0.01249&0.00355&0.00052\\
2&0.58527&0.14903&0.03740&0.01010&0.00072\\
\hline
\end{tabular}
\end{center}
\end{table}

\begin{figure}
\epsfxsize=12cm\epsfbox{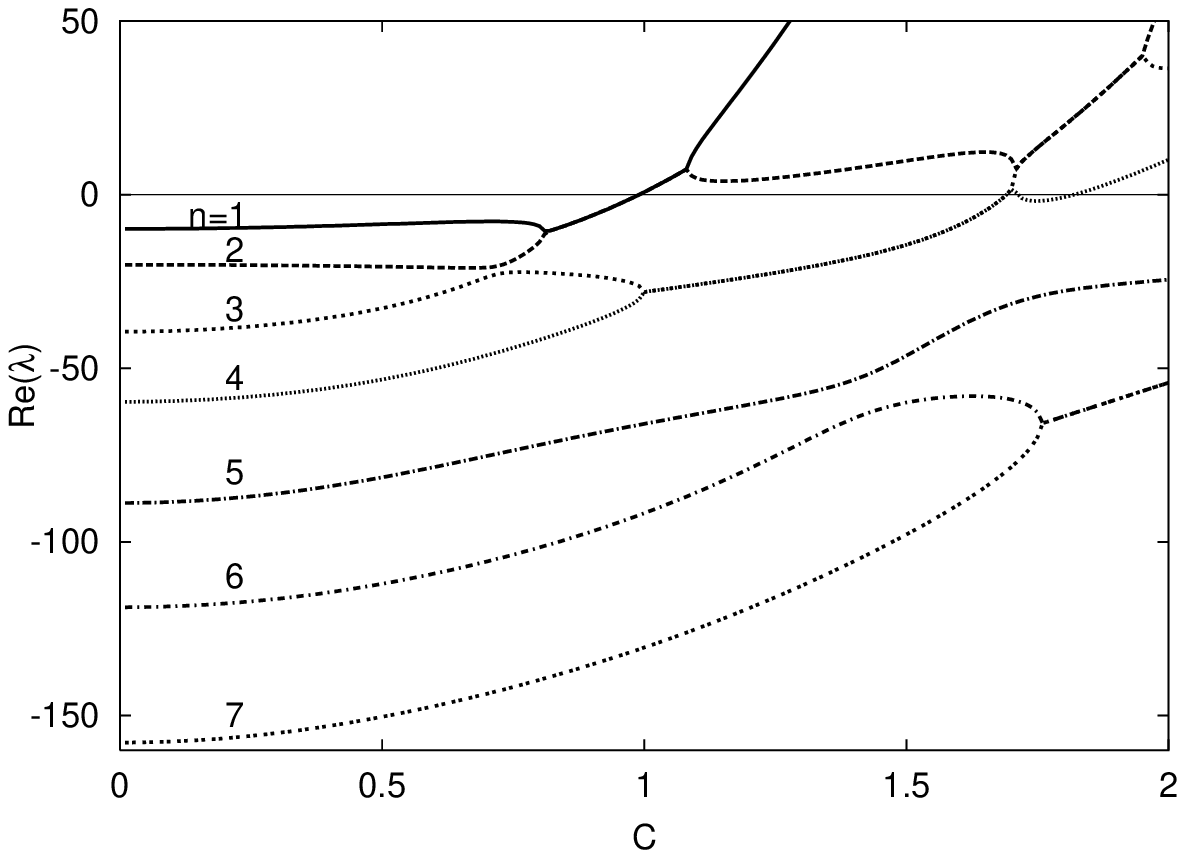}
\epsfxsize=12cm\epsfbox{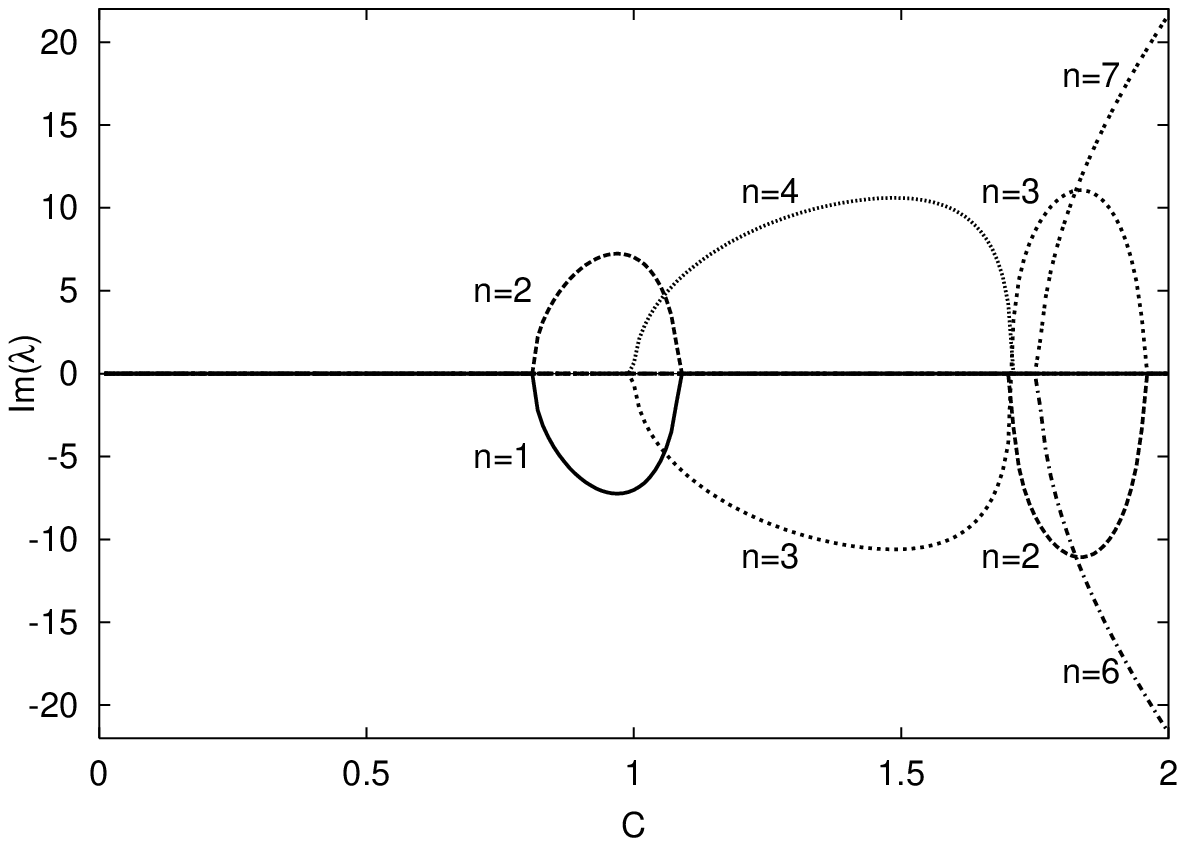}
\caption{Special case $\alpha(x)=C (-21.46+426.41 x^2-806.73 x^3+392.28 x^4$.
Growth rates and frequencies for the eigenfunctions with $(l=1, \; n=1...7)$.
The number of grid points was $N=128$. Note the merging and splitting
points of the spectrum, indicating transitions from non-oscillatory to
oscillatory modes and
vice versa.}
\label{fig4}
\end{figure}

Now, we turn to a more complicated case. It corresponds to the
profile $\alpha(x)=C (-21.46+426.41 x^2-806.73 x^3+392.28 x^4$). 
The choice of this somewhat strange function is motivated by
the fact that it is an example of a proper oscillatory $\alpha^2$ dynamo
\cite{STGE3}. 
In Fig. \ref{fig4} we show the results of the integral equation solver for the case
$l=1$ and  $n=1...7$. We see that the spectral dependence on $C$ 
(Fig. \ref{fig4}) is very complex, with merging and splitting points of neighboring
branches at which non-oscillatory solutions turn into oscillatory solutions and
vice-versa. The computation was done with a grid number of $N=128$, and the
result is basically identical with that of a sophisticated differential
equation solver  \cite{STGE3}. Hence, Fig. \ref{fig4} might serve as a striking example
that the integral equation approach works satisfactorily also in case that
complex eigenvalues appear.

\section{General velocity fields in spherical geometry}

\subsection{Basics}

In this section the Green's function method from Appendix A is 
applied to convert the induction equation for general velocity fields in a unit
sphere   to the integral equations system. In dimensionless form, Eq.
(\ref{eq1}) can be rewritten as follows
\begin{eqnarray}{\label{eqb1}}
\frac{\partial {\bf{B}}}{\partial t}=R_m \nabla\times({\bf{u}}\times 
{\bf{B}})+\nabla^2{\bf{B}},
\end{eqnarray}
where $R_m$ is the magnetic Reynolds number. ${\bf{u}}$ and ${\bf{B}}$ 
may be expanded into the following series (Bullard and Gellman, \cite{BULL}):
\begin{eqnarray}
{\bf{u}}=\sum_\alpha ({\bf{t}}_\alpha+{\bf{s}}_\alpha),{\label{eqb2a}}\\
{\bf{B}}=\sum_\beta ({\bf{T}}_\beta+{\bf{S}}_\beta), {\label{eqb2b}}
\end{eqnarray}
where
\begin{eqnarray}
{\bf{t}}_\alpha&=&\nabla\times [{\bf{e}}_rt_\alpha(r,t)Y_\alpha(\theta,\varphi)],
{\label{eqb3a}}\\
{\bf{s}}_\alpha&=&\nabla\times\nabla\times[{\bf{e}}_rs_\alpha(r,t)
Y_\alpha(\theta,\varphi)],{\label{eqb3b}}
\end{eqnarray}
etc. From here on,  $Y_\alpha$ denote the $(2\alpha+1)$ surface harmonics 
$P_{\alpha m_\alpha}(\theta) sin(m_\alpha\varphi)$, $P_{\alpha m_\alpha}
(\theta)cos(m_\alpha\varphi) (m_\alpha=0, \cdots, \alpha)$, where 
$P_{\alpha m_\alpha}$ is a Legendre function with Neumann normalization, 
$P_{\alpha 0}$ the Legendre polynomial. Similarly $s_\alpha$ is an 
abbreviation for $s_\alpha^{m_\alpha}$, etc. The summations in 
(\ref{eqb2a}) and (\ref{eqb2b}) are over cosine and sine contributions, 
$\alpha=1, 2, 3, \cdots; m_\alpha=0, \cdots, \alpha$; and similarly 
for $\beta, m_\beta$. Substituting  Eqs. (\ref{eqb2a}) and (\ref{eqb2b}) 
into Eq. (\ref{eqb1}), Bullard and Gellman (\cite{BULL}) derived the 
spectral form of (\ref{eqb1}) as
\begin{eqnarray}{\label{eqb4}}
\frac{\partial^2 S_\gamma}{\partial r^2}-\frac{\partial S_\gamma}
{\partial t}-\frac{\gamma(\gamma+1)}{ r^2}S_\gamma=\frac{R_m}{r^2}
\sum_{\alpha, \beta}[(t_\alpha S_\beta S_\gamma)+
(s_\alpha T_\beta S_\gamma)+(s_\alpha S_\beta S_\gamma)]{\label{eqb4a}},\\
\frac{\partial^2 T_\gamma}{\partial r^2}-\frac{\partial T_\gamma}
{\partial t}-\frac{\gamma(\gamma+1)}{r^2}T_\gamma=\frac{R_m}{r^2}
\sum_{\alpha,\beta}[(t_\alpha T_\beta T_\gamma)+(t_\alpha S_\beta 
T_\gamma)+(s_\alpha T_\beta T_\gamma)+(s_\alpha S_\beta T_\gamma)] {\label{eqb4b}},
\end{eqnarray}
where
\begin{eqnarray}{\label{eqb5}}
 (t_\alpha S_\beta S_\gamma)&=&c_1 t_\alpha S_\beta,\nonumber\\
 (s_\alpha T_\beta S_\gamma)&=&c_2 s_\alpha T_\beta, \nonumber\\
(s_\alpha S_\beta S_\gamma)&=&c_3 s_\alpha 
\frac{\partial S_\beta}{\partial r}+c_4\frac{\partial s_\alpha}
{\partial r}S_\beta,\nonumber\\
(t_\alpha T_\beta T_\gamma)&=&c_5 t_\alpha T_\beta, \\
(t_\alpha S_\beta T_\gamma)&=&c_6t_\alpha\frac{\partial S_\beta}
{\partial r}+c_7(\frac{\partial t_\alpha}{\partial r}-
\frac{2t_\alpha}{r})S_\beta, \nonumber\\
(s_\alpha T_\beta T_\gamma)&=&c_8s_\alpha\frac{\partial T_\beta}
{\partial r}+c_8(\frac{\partial s_\alpha}{\partial r}-
\frac{2s_\alpha}{r})T_\beta+c_9\frac{\partial s_\alpha}{\partial r}T_\beta,\nonumber\\
(s_\alpha S_\beta T_\gamma)&=&c_{10}s_\alpha \frac{\partial^2 S_\beta}
{\partial r^2}+c_{11}\frac{\partial s_\alpha}{\partial r}
\frac{\partial S_\beta}{\partial r}+c_{12}\frac{s_\alpha}{r}
\frac{\partial S_\beta}{\partial r}+c_{13}
(\frac{\partial^2 s_\alpha}{\partial r^2}-\frac{2}{r}
\frac{\partial s_\alpha}{\partial r})S_\beta.\nonumber
\end{eqnarray}
The constants $c_i$ in Eq. (\ref{eqb5}) are defined as follows:

\begin{eqnarray}{\label{eqb6}}
c_1&=&-\frac{L\beta(\beta+1)}{N_\gamma}, \nonumber\\
c_2&=&-\frac{L\alpha(\alpha+1)}{N_\gamma},\nonumber\\
c_3&=&-\frac{K}{2N_\gamma}\alpha(\alpha+1)(\alpha(\alpha+1)-
\beta(\beta+1)-\gamma(\gamma+1)),\nonumber\\
c_4&=&-\frac{K}{2N_\gamma}\beta(\beta+1)(\alpha(\alpha+1)-
\beta(\beta+1)+\gamma(\gamma+1)),\nonumber\\
c_5&=&-\frac{L\gamma(\gamma+1)}{N_\gamma},\nonumber\\
c_6&=&-\frac{K}{2N_\gamma}(\beta(\beta+1)(\alpha(\alpha+1)-
\beta(\beta+1)\nonumber\\
&&+\gamma(\gamma+1))+\gamma(\gamma+1)
(\alpha(\alpha+1)+\beta(\beta+1)-\gamma(\gamma+1))),\nonumber\\
c_7&=&-\frac{K}{2N_\gamma}\beta(\beta+1)(\alpha(\alpha+1)-\beta(\beta+1)+
\gamma(\gamma+1)),\\
c_8&=&\frac{K}{2N_\gamma}\alpha(\alpha+1)(-\alpha(\alpha+1)+\beta(\beta+1)+
\gamma(\gamma+1)),\nonumber\\
c_9&=&\frac{K}{2N_\gamma}\gamma(\gamma+1)(\alpha(\alpha+1)+\beta(\beta+1)-
\gamma(\gamma+1)),\nonumber\\
c_{10}&=&\frac{L}{N_\gamma}\alpha(\alpha+1),\nonumber\\
 c_{11}&=&\frac{L}{N_\gamma}(\alpha(\alpha+1)+\beta(\beta+1)-\gamma(\gamma+1)),
\nonumber\\
c_{12}&=&-\frac{2L}{N_\gamma}\alpha(\alpha+1),\nonumber\\
 c_{13}&=&\frac{L}{N_\gamma}\beta(\beta+1).\nonumber
\end{eqnarray}
In Eq. (\ref{eqb6}) we have used the expressions for the Adams-Gaunt and
Elsasser integrals 
\begin{eqnarray}
K&=&\int_0^{2\pi}\int_0^\pi Y_\alpha Y_\beta Y_\gamma sin\theta 
d\theta d\varphi,\nonumber\\
L&=&\int_0^{2\pi}\int_0^\pi Y_\alpha (\frac{\partial Y_\beta}
{\partial \theta}\frac{\partial Y_\gamma}{\partial \varphi}-\frac{\partial Y_\beta}
{\partial\varphi}\frac{\partial Y_\gamma}{\partial \theta}) d\theta d\varphi,
\end{eqnarray}
and the normalization factor
\begin{eqnarray}
N_\gamma&=&\left\{\begin{array}{l}
\frac{2\pi \gamma(\gamma+1)}{2\gamma+1}\frac{(\gamma+m)!}{(\gamma-m)!}, 
m\not=0,\nonumber\\
\frac{4\pi \gamma(\gamma+1)}{2\gamma+1},m=0 .\nonumber\\
\end{array}\right.
\end{eqnarray}

By the same Green's functions method as in Appendix A, and using
the definitions for $G_s(r,r_0)$, $G_t(r,r_0)$ there, the differential
equations system (\ref{eqb4a}) and (\ref{eqb4b})  can be converted to the
following integral equations system: \begin{eqnarray}
S_\gamma(r)&=&\sum_{\alpha \beta}R_m\int_0^1 \frac{G_{s}(r,r_0)}{r_0^2}
[c_1t_\alpha (r_0)S_\beta(r_0)+c_2s_\alpha(r_0)T_\beta(r_0)+
c_3s_\alpha(r_0)\frac{dS_\beta(r_0)}{dr_0}\nonumber\\&&
+c_4\frac{ds_\alpha(r_0)}{dr_0}S_\beta(r_0)]dr_0+\lambda
\int_0^1G_s(r,r_0)S_\gamma(r_0)dr_0, {\label{eqb5a}}\\
T_\gamma(r)&=&\sum_{\alpha \beta}R_m\int_0^1\frac{G_t(r,r_0)}
{r_0^2}[c_5t_\alpha(r_0)T_\beta(r_0)+c_6t_\alpha(r_0)\frac{dS_\beta(r_0)}
{dr_0}+c_7(\frac{dt_\alpha(r_0)}{dr_0}-\frac{2t_\alpha(r_0)}{r_0})
S_\beta(r_0), \nonumber\\
&&+c_8s_\alpha(r_0)\frac{dT_\beta(r_0)}{dr_0}+c_8(\frac{ds_\alpha(r_0)}{dr_0}
-\frac{2}{r_0}s_\alpha(r_0))T_\beta(r_0)+c_9\frac{ds_\alpha(r_0)}{dr_0}
T_\beta(r_0)\nonumber\\
&&+c_{10}s_\alpha(r_0)\frac{d^2S_\beta(r_0)}{dr_0^2}+c_{11}
\frac{ds_\alpha(r_0)}{dr_0}\frac{dS_\beta(r_0)}{dr_0}+c_{12}
\frac{s_\alpha(r_0)}{r_0}\frac{dS_\beta(r_0)}{dr_0}\nonumber\\
&&+c_{13}(\frac{d^2s_\alpha(r_0)}{dr_0^2}-\frac{2}{r_0}
\frac{ds_\alpha(r_0)}{dr_0})S_\beta(r_0)]dr_0+
\lambda\int_0^1G_t(r,r_0)T_\gamma(r_0)dr_0.{\label{eqb5b}}
\end{eqnarray}
Strictly speaking, this is an integro-differential equation system
which could  be used for numerical analysis. If one would insist on
having a pure integral equation system one could employ integration
by parts in order to obtain:
\begin{eqnarray} S_\gamma(r)&=&\sum_{\alpha \beta}R_m[\int_0^1
G_s(r,r_0)F_1(r_0)
S_\beta(r_0)dr_0+\int_0^1G_s(r,r_0)F_2(r_0)T_\beta(r_0)dr_0\nonumber\\
&&+\int_0^1\frac{\partial G_s(r,r_0)} {\partial
r_0}F_3(r_0)S_\beta(r_0)dr_0-\frac{c_3}{2\gamma+1}
r^{\gamma+1}s_\alpha(1.0)S_\beta(1.0)]\nonumber\\
&&+\lambda\int_0^1G_s(r,r_0)S_\gamma(r_0)dr_0,{\label{eqb6a}}\\
T_\gamma(r)&=&\sum_{\alpha\beta}R_m[\int_0^1G_t(r,r_0)F_4(r_0)
T_\beta(r_0)dr_0+\int_0^1G_t(r,r_0)F_5(r_0)S_\beta(r_0)dr_0\nonumber\\
&&+\int_0^1\frac{\partial G_t(r,r_0)}{\partial r_0}F_6(r_0)
S_\beta(r_0)dr_0+\int_0^1\frac{\partial G_t(r,r_0)}{\partial r_0}F_7(r_0)
T_\beta(r_0)dr_0\nonumber\\
&&-c_{10}r^{\gamma+1}s_\alpha(1.0)S_\beta(1.0)+\frac{c_{10}}
{r^2}S_{\alpha}(r)S_\beta(r)]+\lambda\int_0^1G_t(r,r_0)T_\gamma(r_0)dr_0, 
{\label{eqb6b}} \end{eqnarray}  with
\begin{eqnarray}
F_1&=&c_1\frac{t_\alpha}{r_0^2}-c_3\frac{d}{dr_0}(\frac{s_\alpha}{r_0^2})+c_4
\frac{1}{r_0^2}\frac{ds_\alpha}{dr_0},\nonumber\\
F_2&=&c_2\frac{s_\alpha}{r_0^2},\nonumber\\
F_3&=&-c_3\frac{s_\alpha}{r_0^2},\nonumber\\
F_4&=&c_5\frac{t_\alpha}{r_0^2}-c_8\frac{d}{dr_0}(\frac{s_\alpha}{r_0^2})+c_8
\frac{1}{r_0^2}(\frac{ds_\alpha}{dr_0}-\frac{2}{r_0}s_\alpha)+c_9\frac{1}{r_0^2}
\frac{ds_\alpha}{dr_0},\nonumber\\
F_5&=&-c_6\frac{d}{dr_0}(\frac{t_\alpha}{r_0^2})+c_7\frac{1}{r_0^2}
(\frac{dt_\alpha}{dr_0}-\frac{2}{r_0}t_\alpha)+c_{10}
\gamma(\gamma+1)\frac{s_\alpha}{r_0^4}+c_{10}\frac{d^2}
{dr_0^2}(\frac{s_\alpha}{r_0^2})\nonumber\\
&&-c_{11}\frac{d}{dr_0}(\frac{1}{r_0^2}\frac{ds_\alpha}{dr_0})
-c_{12}\frac{d}{dr_0}(\frac{s_\alpha}{r_0^3})+c_{13}\frac{1}
{r_0^2}(\frac{d^2s_\alpha}{dr_0^2}-\frac{2}{r_0}\frac{ds_\alpha}{dr_0}),\nonumber\\
F_6&=&-c_6\frac{t_\alpha}{r_0^2}+2c_{10}\frac{d}{dr_0}(\frac{s_\alpha}{r_0^2})-
c_{11}\frac{1}{r_0^2}\frac{ds_\alpha}{dr_0}-c_{12}\frac{s_\alpha}{r_0^3},\nonumber\\
F_7&=&-c_8\frac{s_\alpha}{r_0^2}\; .\nonumber
\end{eqnarray}

The discretization of this integral equation system is done along the lines
described in subsection \ref{DISCRETIZATION}.

\subsection{A numerical example - The Bullard-Gellman model}

In the following, we will test the suitability of the integral equation
approach to the simulation of large scale velocity fields for a particular 
dynamo model.
In 1954, Bullard and Gellman \cite{BULL} had studied the flow structure
\begin{eqnarray}
{\bf{u}}={\bf{s}}_2^{2c}+5 {\bf{t}}_1^{0} 
\end{eqnarray}
where
\begin{eqnarray}
s_2^{2c}(r)=r^3 (1-r)^2\\
t_1^0(r)=r^2 (1-r)
\end{eqnarray}
claiming that this flow acts indeed as a dynamo.
Later, using higher spatial resolution, Gibson and Roberts \cite{GIRO} and 
Dudley and James \cite{DUJA} falsified
this result showing that there is no dynamo up to a magnetic
Reynolds number of 80.

Here we treat the Bullard-Gellman model within the framework
of the integral equation approach. 

\begin{figure}
\epsfxsize=12cm\epsfbox{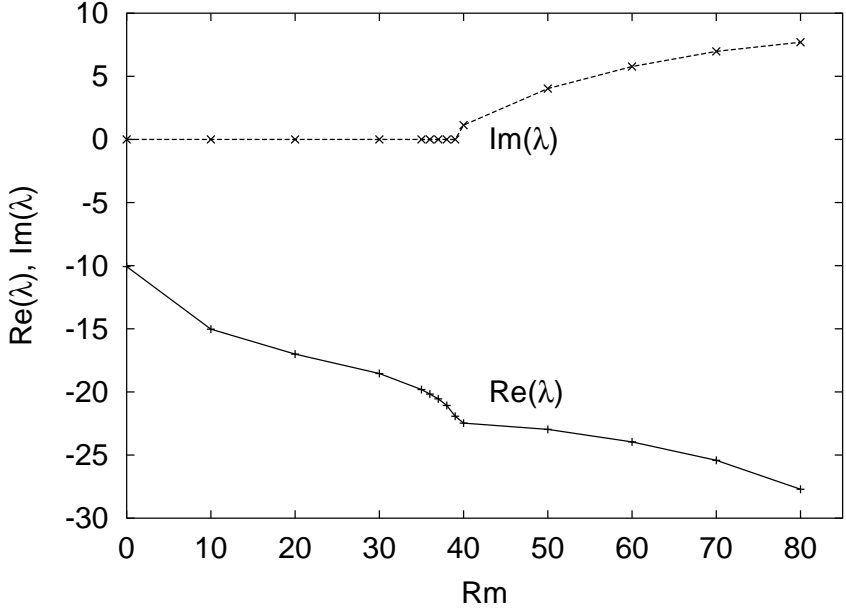}
\caption{Real and imaginary part of the 
first eigenvalue for the Bullard-Gellman dynamo model in dependence
on the magnetic Reynolds number. The truncation degree is $L=9$, the
number of radial grid points is $N=75$.}
\label{fig5}
\end{figure}

In Fig. \ref{fig5} we plot the real and the imaginary part of the
first eigenvalue of the D1 solution (in the terminology of Dudley and
James) for the Bullard-Gellman dynamo model in dependence
on the magnetic Reynolds number. The truncation degree is $L=9$, the
number of radial grid points is $N=75$.
This is essentially the same
curve as published in \cite{DUJA} where a truncation degree $L=12$  and
a number of grid points of $N=100$ had been used, however.

\begin{table}
\caption{Convergence of the integral equation approach (IEA) and the 
Dudley and James method (D\&J). We show the dependence of $\lambda$
on the radial grid number $N$ for $Rm=50$ and $Rm=80$ using a truncation 
degree $L=9$. 
The interpolation in the IEA case is done
with a fit of the date to a function $a+b N^{-2}$. 
Dudley and James had used Richardson
extrapolation based on the values for $N=75,100,125$.}
\label{t4}
\begin{center}
\begin{tabular}{cccccc}\hline
&N=50&N=75&N=100&N=125&Extrapolation\\
\hline
IEA $Rm=50$&-23.50+4.97i&-22.97+4.03i&-22.76+3.66i&-22.63+3.34i\\
IEA $Rm=80$&-29.35+9.46i&-27.70+7.72i&26.95+6.98i&-26.51+6.57i&-26.10+ 6.26i\\
D\&J $Rm=80$&&26.32+6.01i&26.33+6.04i&26.33+6.05i&-26.34+6.06i\\
\hline
\end{tabular}
\end{center}
\end{table}

In Table (\ref{t4}) we have compiled some results concerning
the convergence of our method and the method of Dudley and James.
For $Rm=50$ (where no data are available from Dudley and James)
we see a reasonable convergence of the real part but a slow convergence of the
imaginary part. The latter might be due to the fact that we are not far
from the transition point to oscillatory behaviour where the imaginary 
part is sensible to changes in the grid number.
For $Rm=80$ we have to  concede that the convergence in 
our case is slower than in
the differential
equation method of Dudley and James.

A similar conclusion can be drawn from the treatment of other models
(Lilley model, modified Lilley model).
Although our method yields essentially the same results as the
differential equation approach, it seems worth to look for refined numerical
methods to solve the integral eigenvalue  equation.

\section{Remarks and conclusions}

We have established the integral equation approach to time-dependent kinematic
dynamos in arbitrary domains. This approach is based on Biot-Savart's law. 
The main advantage of the method is its suitability to handle 
dynamos in arbitrary domains. The necessity to solve the Laplace 
equation
in the exteriour of the dynamo domain is circumvented by the 
(implicit) solution of boundary integral equations for the electric potential
and the magnetic vector potential.

It should be noted that we have worked out only one possible form of the
integral equation approach which results in a linear eigenvalue problem.
Another form could start from rewriting Eq. (\ref{eq12}) into the form
of a Helmholtz equation for the vector potential
\begin{eqnarray}
\Delta {\bf{A}}-\mu_0 \sigma \lambda {\bf{A}}=\mu_0 \sigma({\bf F}-\nabla \varphi) \; .
\end{eqnarray}
Then, the pendant to Eq. (\ref{eq13}) would read 
\begin{eqnarray}
{\bf{B}}({\bf{r}})&=&\frac{\mu_0 \sigma}{4 \pi} \int_D
\frac{ {\bf{F}}({\bf{r'}}) \times 
({\bf{r}}-{\bf{r'}})}{|{\bf{r}}-{\bf{r'}}|^3} 
\exp{(k |{\bf{r}}-{\bf{r'}}|)} (1-k |{\bf{r}}-{\bf{r'}}|)\; dV'  \nonumber\\
&&
-\frac{\mu_0 \sigma}{4 \pi} \int_S \varphi({\bf{s'}}) 
{\bf{n}} ({\bf{s'}}) \times
\frac{{\bf{r}}-{\bf{s'}}}{|{\bf{r}}-{\bf{s'}}|^3} 
\exp{(k |{\bf{r}}-{\bf{r'}}|)} (1-k |{\bf{r}}-{\bf{r'}}|)\;
\; dS',
\end{eqnarray}
with 
\begin{eqnarray}
k=\sqrt{\lambda \mu_0 \sigma} \; .
\end{eqnarray}
Without going into the details of such a formulation (we skip here
the equations for the electric potential and the vector potential at
the boundary), we see immediately that we end up with a 
non-linear eigenvalue equation for the eigenvalue $\lambda$. 
It would be interesting to compare the numerical performance of such a formulation 
with the present one.

We plan to use our formulation for a number of dynamo problems, in 
particular problems
which are connected with the design and optimization for new dynamo experiments and
with velocity reconstruction problem for the existing experiments. 

\section{Appendix A}
In this part, we give another approach to establish the integral 
equations (\ref{eq27}) and (\ref{eq41}). We start with the following 
differential equation problem:
\begin{eqnarray}
\lambda s_{lm}&=&\frac{1}{\mu_0\sigma}[\frac{d^2s_{lm}}{dr^2}-
\frac{l(l+1)}{r^2}s_{lm}]+\alpha(r)t_{lm},\label{eq59}\\
\lambda t_{lm}&=&\frac{1}{\mu_0\sigma}[\frac{d^2t_{lm}}{dr^2}-
\frac{l(l+1)}{r^2}t_{lm}]-\frac{d}{dr}(\alpha(r)\frac{ds_{lm}}{dr})+
\frac{l(l+1)}{r^2}\alpha(r)s_{lm},\label{eq60}\\
t_{lm}(R)&=&R\frac{ds_{lm}}{dr}|_{r=R}+l s_{lm}(R)=0,\label{eq61}\\
\nonumber
\end{eqnarray}
First, we derive the Green's function, $G_s(r,r_0)$,
corresponding to the equation (\ref{eq59}). This Green's function satisfies:
\begin{eqnarray} \frac{\partial^2G_s}{\partial
r^2}-\frac{l(l+1)}{r^2}G_s&=&\delta(r-r_0),\label{eq62}\\
G_s|_{r=0}&=&0,\label{eq63}\\ 
R \frac{\partial G_s}{\partial r}|_{r=R}+l G_s|_{r=R}&=&0 \; .\label{eq64}
\end{eqnarray}
According to the construction method of Green's functions (\cite{COUR}, P.
355), we obtain $G_s(r,r_0)$ in the following form:
\begin{eqnarray}{\label{eq66}} G_s(r,r_0)=\left\{\begin{array}{l}
-\frac{1}{2l+1}r_0^{-l}r^{l+1},r\leq r_0,\\
-\frac{1}{2l+1}r_0^{l+1}r^{-l},r\geq r_0.\\
\end{array}\right.
\end{eqnarray}
As for Eq. (\ref{eq60}), we first rewrite it in the form:
\begin{eqnarray}{\label{eq67}}
\frac{1}{\mu_0 \sigma}\frac{d^2F}{dr^2}-\frac{1}
{\mu_0\sigma}\frac{l(l+1)}{r^2}F+\frac{d}{dr}(\frac{d\alpha}{dr}s_{lm})-
\mu_0\sigma\lambda t_{lm}=0,
\end{eqnarray}
where $F=t_{lm}-\mu_0\sigma\alpha s_{lm}$.
This differential equation problem for $t_{lm}$ can be split into two
problems. One of them reads
\begin{eqnarray}{\label{eq68}}
\frac{d^2F_1}{dr^2}-\frac{l(l+1)}{r^2}F_1+\mu_0\sigma\frac{d}{dr}(s_{lm}
\frac{d\alpha}{dr})-\mu_0\sigma\lambda t_{lm}=0,\\
F_1|_{r=R}=0.\nonumber
\end{eqnarray}
The other is
\begin{eqnarray}{\label{eq69}}
\frac{d^2F_2}{dr^2}-\frac{l(l+1)}{r^2}F_2=0,\\
F_2|_{r=R}=-\mu_0\sigma\alpha(R)s_{lm}(R).\nonumber
\end{eqnarray}
For the differential equation problem (\ref{eq68}), applying the construction 
method of Green's function (\cite{COUR}) again, we obtain its Green's
function in the form: \begin{eqnarray}{\label{eq70}}
G_t(r,r_0)=\left\{\begin{array}{l}
\frac{1}{R^{2l+1}(2l+1)}(r_0^{l+1}-R^{2l+1}r_0^{-l})r^{l+1}, r\leq r_0,\\
\frac{1}{R^{2l+1}(2l+1)}(r^{l+1}-R^{2l+1}r^{-l})r_0^{l+1}, r\geq r_0\; ,
\end{array}\right.
\end{eqnarray}
which satisfies
\begin{eqnarray} 
\frac{\partial^2G_t}{\partial
r^2}-\frac{l(l+1)}{r^2}G_t&=&0\\
G_t|_{r=0}&=&0,\\ 
G_t|_{r=R}&=&0   \; .
\end{eqnarray}

As for the differential equation problem (\ref{eq69}), the solution can 
be expressed as: \begin{eqnarray}{\label{eq71}}
F_2=-\frac{\mu_0\sigma}{R^{l+1}}\alpha(R)s_{lm}(R)r^{l+1}.
\end{eqnarray}
Then the superposition theorem of the linear problems 
allows us to obtain the following integral equations for $s_{lm}$ and $t_{lm}$:
\begin{eqnarray}
s_{lm}(r)&=&-\int_0^RG_s(r,r_0)\mu_0\sigma\alpha(r_0)t_{lm}(r_0)dr_0+\mu_0\sigma\lambda \int_0^R G_s(r,r_0)s_{lm}(r_0)dr_0,\label{eq72}\\
t_{lm}(r)&=&\mu_0\sigma\alpha(r)s_{lm}(r)-\int_0^R\mu_0\sigma\frac{d}{dr_0}(s_{lm}(r_0)\frac{d\alpha(r_0)}{dr_0})G_t(r,r_0)dr_0+\nonumber\\&&\lambda\mu_0\sigma\int_0^RG_t(r,r_0)t_{lm}(r_0)dr_0
-\frac{r^{l+1}}{R^{l+1}}\mu_0\sigma\alpha(R)s_{lm}(R).\label{eq73}
\end{eqnarray}
Integrating by parts the terms containing 
derivatives of $s_{lm}$ in Eq. (\ref{eq73}), we obtain
\begin{eqnarray}
s_{lm}(r)&=&-\int_0^RG_s(r,r_0)\mu_0\sigma\alpha(r_0)t_{lm}(r_0)dr_0+\mu_0\sigma\lambda \int_0^R G_s(r,r_0)s_{lm}(r_0)dr_0,\label{eq74}\\
t_{lm}(r)&=&\mu_0\sigma\alpha(r)s_{lm}(r)+\int_0^R\mu_0\sigma\frac{d\alpha(r_0)}{dr_0}s_{lm}(r_0)\frac{\partial G_t(r,r_0)}{\partial r_0}dr_0\nonumber\\&&+\lambda\mu_0\sigma\int_0^RG_t(r,r_0)t_{lm}(r_0)dr_0
-\frac{r^{l+1}}{R^{l+1}}\mu_0\sigma\alpha(R)s_{lm}(R).\label{eq74}
\end{eqnarray}
Therefore, we have obtained the same integral equations as expressed in 
Eqs.
(\ref{eq27}) and (\ref{eq41}).

\section*{ACKNOWLEDGMENTS}

Financial support from German "Deutsche Forschungsgemeinschaft" in frame
of the Collaborative Research Center SFB 609 is gratefully acknowledged.

\end{document}